\documentclass[12pt]{article}
\usepackage{epsfig}
\usepackage{fullpage}
\def\vev#1{\left\langle #1 \right\rangle}

\long\def\symbolfootnote[#1]#2{\begingroup%
\def\thefootnote{\fnsymbol{footnote}}\footnote[#1]{#2}\endgroup}

\newcommand{\be}{\begin{equation}}
\newcommand{\ee}{\end{equation}}
\newcommand{\bea}{\begin{eqnarray}}
\newcommand{\eea}{\end{eqnarray}}
\newcommand{\non}{\nonumber}
\newcommand{\sst}{\scriptstyle}
\newcommand{\tn}{\textnormal}

\def\epl#1{\varepsilon_{#1}}
\def\epu#1{\varepsilon^{#1}}
\def\De#1{ \delta^{(3)}(#1) }
\def\de#1{ {\sst \delta^{(3)}(#1) } }

\def\q#1{q_{#1}}

\def\al#1{\alpha_{#1}}
\def\bl#1{\beta_{#1}}
\def\gl#1{\gamma_{#1}}
\def\m#1{\mu_{#1}}
\def\kt#1{ \tilde{k}^{\mu_{#1}}}
\def\kpt#1{ \tilde{k}^{\prime\mu_{#1}}}
\def\rh#1{\rho_{#1}}
\def\e#1{\textnormal{e}^{#1}}
\def\kp{k^{\prime}}
\def\si#1#2{ {\sst \sin{(\frac{#1\times#2}{2})} } }
\def\bisi{ {\sst \sin{(\frac{\eta\times q - k\times \eta-k\times
q}{2})} } }
\def\propone{ {\sst \frac{ k^{\rh1} (k-\eta)^{\rh2} \eta^{\rh3} q^{\rh4}
  (q+\eta)^{\rh5} }{k^2 (k-\eta)^2 \eta^2 q^2 (q+\eta)^2} } } 

\def\proptwo{ {\sst \frac{ (k-\eta)^{\rh2} \eta^{\rh3} q^{\rh4}
  (q+\eta)^{\rh5} (k+q)^{\rh6}}{(k-\eta)^2 \eta^2 q^2 (q+\eta)^2 (k+q)^2}} }

\def\propthree{ {\sst \frac{ k^{\rh1} \eta^{\rh3} q^{\rh4}
(q+\eta)^{\rh5} (k+q)^{\rh6}} { k^2 \eta^2 q^2 (q+\eta)^2 (k+q)^2}} }

\def\propfour{ {\sst \frac{ k^{\rh1} (k-\eta)^{\rh2} \eta^{\rh3} q^{\rh4}
  (q+\eta)^{\rh5} (k+q)^{\rh6} }{k^2 (k-\eta)^2 \eta^2 q^2 (q+\eta)^2
  (k+q)^2 } } }
 
\def\itwo{ \frac{i}{2} }

\begin{document}
\baselineskip=15.5pt
\renewcommand{\theequation}{\arabic{section}.\arabic{equation}}
\setcounter{page}{1}

%--------+---------+---------+---------+---------+---------+---------+
%Title page
\begin{titlepage}

\leftline{\tt hep-th/0310011}

\vskip -.5cm

\rightline{\small{\tt CALT-68-2453}}

\begin{center}

\vskip 2 cm

{\Large {On the quantum equivalence of commutative and}}

\vskip .5cm

{\Large{noncommutative Chern-Simons theories at higher orders}}

\vskip 1.5cm
{\large Kirk Kaminsky\symbolfootnote[2]{
e-mail address:\ \ kaminsky@theory.caltech.edu}}
\vskip 1.0cm

{{\it California Institute of Technology}}

\smallskip

{{\it Mail Code 452-48}}

\smallskip

{{\it Pasadena, CA 91125 USA}}

\vskip 2cm

{\bf Abstract}
\end{center}

\noindent
We continue our investigation of the quantum equivalence between
commutative and noncommutative Chern-Simons theories by computing
the complete set of two-loop quantum corrections to the correlation
function of a pure open Wilson line and an open Wilson line with 
a field strength insertion, on the noncommutative side in a covariant
gauge.  The conjectured perturbative equivalence between the free commutative
theory and the apparently interacting noncommutative one requires
that the sum of these corrections vanish, and herein we exhibit the remarkable
cancellations that enforce this.  From this computation we speculate
on the form of a possible all-order result for this simplest
nonvanishing correlator of gauge invariant observables.

\end{titlepage}

\newpage

%--------+---------+---------+---------+---------+---------+---------+
%Body

%%%%% Section 1 %%%%%
\section{Introduction}
\setcounter{equation}{0}

In \cite{Grandi:2000, Polychronakos:2002} it was shown that the
Chern-Simons action on a noncommutative spacetime given by
\be
S_{NCCS} = \frac{1}{2} \int d^3 x~ \epu{\mu \rho \nu}
  \left[
    A_\mu \ast \partial_\rho A_\nu
    - \frac{2ig}{3} A_\mu \ast A_\rho \ast A_\nu
  \right],
\label{NCCS}
\ee
with the standard star product
\be
f(x)\ast g(x) \equiv \e{\frac{i}{2}\theta^{\mu\nu} \partial^y_\mu
\partial^z_\nu} f(y)g(z) \bigg|_{y,z\rightarrow 0}
\ee
is a fixed-point of the Seiberg-Witten map\cite{Seiberg:1999}: 
\bea
\delta A^{\mu}&=& \delta \theta^{\rho\sigma} \frac{\partial}
{\partial\theta^{\rho\sigma}} A = -\frac{1}{4} \delta
\theta^{\rho\sigma} \left\{ A_{\rho}, \partial_{\sigma}
 A_{\mu} + F_{\sigma\mu} \right\}_\ast \non \\ 
\delta F^{\mu\nu} &=& \delta^{\rho\sigma}\frac{\partial}
{\partial\theta^{\rho\sigma}} F_{\mu\nu} = \frac{1}{4} \delta
\theta^{\rho\sigma} \left( 2 \left\{
F_{\mu\rho},F_{\nu\sigma} \right\}_\ast - \left\{
A_{\rho} , D_{\sigma} F_{\mu\nu} + \partial_{\sigma}
F_{\mu\nu} \right\}_\ast \right).
\eea
Here $D_{\mu} \psi = \partial_{\mu} \psi - ig
[A_{\mu}, \psi]_{\ast}$, and $F_{\mu\nu} =
\partial_{\mu}A_{\nu} -\partial_{\nu}A_{\mu}-ig[ A_{\mu}
, A_{\nu}]_\ast$.  In particular
\be
\frac{\partial S_{NCCS}(A)}{\partial \theta^{\rho\sigma}}
= 0.
\ee
This implies that noncommutative and commutative Chern-Simons theories
are classically equivalent under the Seiberg-Witten map.\footnote{This
map derives from a {\it disk} computation in string
theory, and is therefore itself a {\it classical} map.}  This is
important for two reasons.  First, in contradistinction to the usual
scenario, it identifies an instance where the action on the
commutative side of the map is known in
closed form.  Second, because the correlation functions of
observables in  that theory can therefore be
computed explicitly (and in the $U(1)$ case exactly, in a covariant
gauge), we can test the equivalence of the theories at the
quantum level by computing the correlation functions of their
Seiberg-Witten transforms.\footnote{It also gives rise to the
puzzle where in the $U(1)$ case the commutative space action is free,
whereas the noncommutative space action is an apparently interacting theory
with a nontrivial coupling constant.}  In turn this first requires us
to identify what observables need to be computed on the
noncommutative side.

It is well-known \cite{Ishibashi:1999, Das:2000} that there are
no locally gauge-invariant observables
in position space in noncommutative gauge theories because gauge 
transformations act as spacetime translations, as a consequence of the
basic noncommutative identity
\be
\e{i k\cdot x} \ast U(x+ k\theta) = U(x) \ast \e{i k\cdot x}.
\label{noncommutative identity}
\ee
However, it was shown in \cite{Das:2000, Gross:2000} that a
complete set of gauge-invariant observables in noncommutative gauge
theories which are local in {\it momentum} space are provided by the
Fourier transform of open Wilson lines:
\bea
W(k)&=& \int d^3x~ P_{\ast}\exp{\left[ig \int_0^1 d\sigma \kt{} 
A_{\mu}(x+\xi(\sigma))\right]} \ast \e{ik\cdot x}, \\
O(k) &=& \int d^3x~ P_{\ast}\exp{\left[ig \int_0^1 d\sigma \kt{} 
A_{\mu}(x+\xi(\sigma))\right]} \ast O(x) \ast \e{ik\cdot x},
\eea
where $O(x)$ is any local operator transforming in the adjoint
representation.
Furthermore, a closed form of the Seiberg-Witten map for the case of a
$U(1)$ gauge group  was exhibited in
\cite{Okawa:2001, Mukhi:2001, Liu:2001}, which reduces in three
dimensions to
\bea
 f_{12}(k) &=& - \frac{1}{g\theta^{12}}\left[W(k) -
(2\pi)^3\de{k}\right] \non \\
 f_{0i}(k) &=& O_{0i}(k),
\eea
in a coordinate system where only  $\theta^{12}= -\theta^{21}$ is
nonvanishing. Here $f_{\mu\nu}(k)$ is the commutative field strength
in momentum space, and $O_{\mu\nu}(k)$ is the open Wilson line with a
noncommutative field strength insertion at one end:
\be
O_{\mu\nu}(k) = \int d^3x~ P_{\ast}\exp{\left[ig \int_0^1 d\sigma \kt{} 
A_{\mu}(x+\xi(\sigma))\right]} \ast F_{\mu\nu}(x) \ast \e{ik\cdot x}.
\ee
This provides the required correspondence between observables on the
commutative and noncommutative sides, and by computing their
correlators, allows us to test the quantum equivalence of the
two theories.

On the commutative side, the simplest nonvanishing correlator of gauge
invariant observables in $U(1)$ Chern Simons theory\footnote{We will
focus on the $U(1)$
case in this paper which is of interest in part due 
to its conjectured relevance in the microscopic description fractional
quantum Hall fluids \cite{Susskind:2001,Polychronakos:2001,Hellerman:2001rj, 
Bergman:2001,Hellerman:2001yv}.
The generalization to the $U(N)$ case, which
amounts to homogeneous factors of powers of $N$, is straightforward and
was discussed by us in \cite{Kaminsky:2003}.} is the two-point
function of $f_{12}$ and $f_{0i}$, and is given exactly in momentum space
by
\be
\vev{f_{12}(k)f_{0i}(k^{\prime})} = (2\pi)^3 k_i \delta^{(3)}(k+k^{\prime}).
\label{commutative correlator}
\ee 
Noncommutative $U(1)$ Chern-Simons theory contains an apparently
nontrivial interaction with a coupling constant $g$, and thus the
computation on the noncommutative side is at the outset
nontrivial in a covariant gauge.  This is further complicated by the
fact that on the
noncommutative side, we have to compute the correlator of composite
objects, which even in free commutative theories can be nontrivial.
In \cite{Kaminsky:2003} we computed the Seiberg-Witten transform of
(\ref{commutative correlator}), $\vev{W(k)O_{\mu\nu}(k^{\prime})}$, to
$O(g^3)$ in the gauge coupling $g$ (i.e. one-loop), and found that the
$O(g)$ term reproduced the commutative result while the one-loop or
$O(g^3)$ contributions either cancelled amongst themselves, or yielded
a harmless wavefunction renormalization to the Seiberg-Witten map
itself.  Herein we proceed to compute the complete set of two-loop,
$O(g^5)$ contributions to this correlator with the expectation that it
receives no quantum correction at this order, as required by the
conjectured equivalence between the two theories.  From this
computation, and that done previously in
\cite{Kaminsky:2003}, we will hopefully acquire enough insight to
conjecture the complete cancellation of quantum corrections to all
orders in perturbation theory.

%%%%% Section 2  %%%%% 
\section{Setup}
\setcounter{equation}{0}

In the interests of maintaining maximum transparency in this lengthy
calculation and emphasizing the
essentially algebraic nature of the cancellations present, we will work
formally and defer discussion of regularization to the end.
However we will assume that the point-splitting regulator introduced in
\cite{Kaminsky:2003}, and its obvious generalization to handle the
inclusion of three gauge field sources, or the field strength
commutator and two gauge field sources can be applied: we
assume that all operators on $W(k)$ and $O_{\mu\nu}(k^{\prime})$ are
separated by a path parameter spacing $\epsilon$, so as to ensure that
a nonvanishing noncommutative phase is always present to regulate the
diagrams with respect to the associated loop integration.  For the
graphs we need to explicitly evaluate that involve 'internal' loops,
there are also logarithmic divergences coming from the planar parts
of those loops that are of the same form found in commutative
Chern-Simons theories, and which we will assume are regulated in an
appropriate manner.  

We will also invoke several notational simplifications to
prevent the expressions from becoming too unwieldy and to hopefully make the
calculation easier to follow.  To this end we
will absorb the homogeneous factor of $(2\pi)^{-3}$, the overall
momentum conservation delta function $\De{k+\kp}$ as well as the 
delta functions expressing momentum conservation at
the vertices, and the integration measures
into a generalized integration symbol when there is no ambiguity,
with the understanding that all momenta except for $k$, the momentum
carried by $W(k)$, and all path
parameters are to be integrated over their appropriate ranges.
Spacetime indices and
(initially) momenta will be labelled according to their origin from
within the expansion of $W(k)$ or $O_{\mu\nu}(\kp)$.  Thus $p_{ij}$ refers
to the momentum carried by the $i$th gauge field from the $j$th Wilson line.
Specifically, $p_{i1}$ will refer to
the momenta of gauge fields on $W(k)$, $p_{i2}$ to those from
$O_{\mu\nu}(\kp)$, and since we will not require more than two gauge
field sources from the path-ordered exponential
part of $O_{\mu\nu}(\kp)$, the field
strength commutator momenta will always be labelled by $p_{32}$ and
$p_{42}$.  Similarly the indices $\m{ij}$ will always be associated
with the basic line element $\tilde{k} \equiv k\theta$ (and are hence 
interchangeable within tensor expressions), while indices 
$\alpha_i,\beta_i,\gamma_i$ and respective momenta $q_i,r_i,s_i$ are 
associated with contractions into internal vertices.  

We will subsequently re-define or re-label the momenta to a standard
set, which will make the momentum constraints manifest, homogenize
the expressions, and allow us to compare diagrams by inspection of the
resulting tensor expressions.  Thus, throughout the computation the 
indices $\rh{1}, \rh{2}, \ldots \rh{7}$ will always be associated with
momenta $k, k-\eta, \eta, q, q+\eta, k+q,$ and $k$ respectively,
where $\eta$ and $q$ will be our loop integration variables.  The
calculation will be presented without recourse to expanding pairs of
antisymmetric symbols lacking mutual contracted indices by employing
identities presented in the appendix to write all final tensor
expressions in terms of $\epl{\mu\nu a}$, where $a$ is any of
$\rh{1}...\rh{7}$, or $\m{ij}$.  The only unsummed indices are
$\mu$,$\nu$ and the only unintegrated momentum is $k$.  

We will freely
use the fact that in Chern-Simons theory a direct contraction between any
pair of gauge fields from the Wilson lines themselves vanish in the
Landau gauge, as do both one-point
tadpoles and contributions where the two Wilson lines are
disconnected.  Additionally, since
the possible one-loop correction to the Chern-Simons propagator changes neither
its tensor structure nor its momentum dependence as discussed in
\cite{Kaminsky:2003}, the {\it two}-loop contributions involving it are
necessarily a repeat of the {\it one}-loop calculation presented
therein, and we will henceforth ignore them.\footnote{Alternatively,
as is well-known, the gauge and ghost graphs involved in the one-loop
propagator correction formally cancel, and the one-loop shift arises
from the regulator, so
each of these putative contributions at two-loops pairwise cancel at
the formal level.}  These observations allow us to drastically reduce
the number of diagrams and contractions we must consider.

A key {\it a posteriori} insight acquired from this calculation is to collect
the contributions according to the number of gauge field sources on
the pure open Wilson line $W(k)$, a fact that will be the basis for
our all-orders discussion at the end.  In any case, the set of cancellations we
now exhibit at two-loops will be presented in this way.  We label all
diagrams as $[x.yy]$, where $x$ denotes the number of sources on $W(k)$, and
$yy$ is a counter.

%%%%%  Section 3  %%%%%
\section{Contributions from $O(g^3)$ terms in $W(k)$}
\setcounter{equation}{0}

%%%%% Figure 1 %%%%%
\begin{figure}
\centerline{\epsfxsize=2in\epsfbox{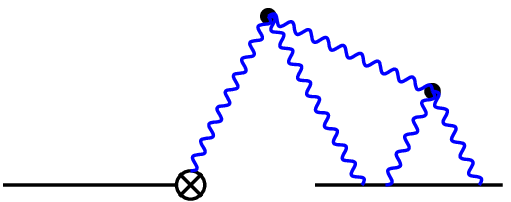} \hspace{1cm}
\epsfxsize=2in\epsfbox{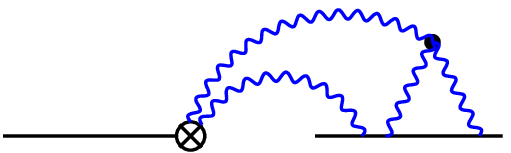}}
\caption{One of three pairs of cancelling graphs, [3.01] and [3.02], with
three gauge field sources on the pure open Wilson line.}
\end{figure}
%%%%%%%%%%%%%%%%%%%%

Figure 1 illustrates one of three pairs of diagrams that involve
contractions involving three gauge field sources from $W(k)$ and no
gauge field sources from the path-ordered exponential component of 
$O_{\mu\nu}(k^{\prime})$.  Since we will not need to perform the path
parameter or the loop integrations, it suffices to consider one pair;
the other two cases are identical.

For the graph with two internal vertices we have
\bea
[3.01]&=& (ig)^3(-2ig)^2\int \kt{11}\kt{21}\kt{31} \de{ p_{11}
+ p_{21} + p_{31}-k} \de{p_{42}-\kp} \e{-i\left[p_{11}\cdot \xi_{11} +
p_{21}\cdot \xi_{21} + p_{31}\cdot \xi_{31}\right]} \non \\
& & ~\times \e{-\frac{i}{2}\left[ p_{21}\times
(p_{31}-k) -p_{31}\times k\right]} \si{q_1}{q_2} \si{r_1}{r_2}
\frac{p_{42}^{\rh1} p_{11}^{\rh2} q_3^{\rh3} p_{21}^{\rh4}
p_{31}^{\rh5}}{p_{42}^2 p_{11}^2 q_3^2 p_{21}^2 p_{31}^2} \non \\
& & ~ \times (-ip_{42})_{\mu}
\epl{\nu \rh{1}\al{1}} \epl{\m{11}\rh{2}\al{2}} 
\epl{\al{3}\rh{3}\bl{1}}
\epl{\m{21}\rh{4}\bl{2}} \epl{\m{31}\rh{5}\bl{3}} \epu{\al1 \al2 \al3}
\epu{\bl1 \bl2 \bl3} \non \\
& & ~ \times \de{p_{42}+q_1} \de{p_{11}+q_2}\de{q_3+r_1}
\de{p_{21}+r_2} \de{p_{31}+r_3} ~~ - ~~ (\mu \leftrightarrow \nu).
\eea

To satisfy the momentum constraints define $p_{11}= k-\eta$, $p_{21}=
-q$, $p_{31}= q+\eta$, $p_{42}= -k$, and $q_3 = -\eta$.  Then
contracting on $\al2, \bl3$ using (\ref{epsilon identity 1}) we obtain
\bea
[3.01]&=& -4 g^5 \int \kt{11}\kt{21}\kt{31} \e{ {\sst 
i[(k\times\eta)\sigma_{11}
+ (k\times q)\sigma_{21} - k\times(q+\eta) \sigma_{31}] } } 
\e{ -\frac{i}{2} (\eta\times q + k\times \eta)} \non \\
& & \si{k}{\eta}\si{\eta}{q} \propone ~ 
\epl{\m{31}\rh{2}\rh{3}}\epl{\m{21}\rh{4}\rh{5}} k_{ {\sst[} \mu}
\epl{ \nu {\sst ]}\rh{1}\m{11}} \non \\
&=& 4 g^5 \int \kt{11}\kt{21}\kt{31} 
\e{ {\sst i[(k\times\eta)\sigma_{11}
+ (k\times q)\sigma_{21} - k\times(q+\eta) \sigma_{31}] } } 
\e{ -\frac{i}{2} (\eta\times q + k\times \eta)} \non \\ 
& & \si{k}{\eta}\si{\eta}{q}
{\sst \frac{ (k-\eta)^{\rh2} \eta^{\rh3} q^{\rh4}
  (q+\eta)^{\rh5} }{(k-\eta)^2 \eta^2 q^2 (q+\eta)^2} } 
\epl{\m{31}\rh{2}\rh{3}}\epl{\m{21}\rh{4}\rh{5}} \epl{\mu\nu\m{11}}.
\non \\
\eea
In the last line we have used (\ref{master identity}) with $C=k$.

There are two contractions into the commutator in the second graph:
\bea
[3.02]&=& -(ig)^4(-2ig)\int \kt{11}\kt{21}\kt{31}
 \de{ p_{11}+ p_{21} + p_{31}-k} \de{p_{32}+ p_{42}-\kp} 
\e{-i\left[p_{11}\cdot \xi_{11} +
p_{21}\cdot \xi_{21} + p_{31}\cdot \xi_{31}\right]} \non \\
& & ~\times \e{-\frac{i}{2}\left[ p_{21}\times
(p_{31}-k) -p_{31}\times k\right]} \e{-\frac{i}{2} p_{32} \times 
p_{42} } \si{r_1}{r_2} \epu{\bl1 \bl2 \bl3}
\frac{ p_{11}^{\rh2} p_{21}^{\rh4}
p_{31}^{\rh5}}{p_{11}^2 p_{21}^2 p_{31}^2}  \non \\
& & \times \left\{ \epl{\m{11}\rh{2} \mu } 
\epl{\nu \rh{3}\bl{1}} \epl{ \m{21} \rh{4} \bl{2}} \epl{\m{31} \rh{5} \bl{3}}
{\sst \frac{p_{42}^{\rh{3}}}{p_{42}^2} }
\de{p_{11}+ p_{32}} \de{p_{21}+r_2}\de{p_{31}+r_3}\de{p_{42}+r_1}~ +
\right. \non \\
& & \left. ~ \epl{\m{11}\rh{2}\nu} \epl{\mu\rh{3}\bl{1}} \epl{ \m{21}
\rh{4} \bl{2}} \epl{\m{31} \rh{5} \bl{3}} 
{\sst \frac{p_{32}^{\rh{3}}}{p_{32}^2} }
\de{p_{11}+ p_{42}} \de{p_{21}+r_2}\de{p_{31}+r_3}\de{p_{32}+r_1}
\right\} - {\sst (\mu \leftrightarrow \nu)}. \non \\
\eea
Again define $p_{11}=k-\eta$, $p_{21}=-q$, $p_{31}=q+\eta$.  Then
contracting on $\bl3$ and using (\ref{epsilon identity 4}) we obtain
\bea
[3.02]&=& 2ig^5 \int \kt{11}\kt{21}\kt{31}
\e{ {\sst i[(k\times\eta)\sigma_{11}
+ (k\times q)\sigma_{21} - k\times(q+\eta) \sigma_{31}] } } 
\e{ -\frac{i}{2} (\eta\times q + k\times \eta)} 
{\sst \frac{ (k-\eta)^{\rh2} \eta^{\rh3} q^{\rh4}
  (q+\eta)^{\rh5} }{(k-\eta)^2 \eta^2 q^2 (q+\eta)^2} }
\non \\
& & \si{\eta}{q}
\left\{ \epl{\m{11}\rh{2}\mu} \epl{\nu\rh{3}\m{31}}
\epl{\m{21}\rh{4}\rh{5}} \e{-\frac{i}{2}k\times\eta} +  
\epl{\m{11}\rh{2}\nu} \epl{\mu\rh{3}\m{31}}
\epl{\m{21}\rh{4}\rh{5}} \e{\frac{i}{2}k\times\eta} 
 \right\} - (\mu \leftrightarrow \nu) \non \\
&=& -4g^5 \int \kt{11}\kt{21}\kt{31}
\e{ {\sst i[(k\times\eta)\sigma_{11}
+ (k\times q)\sigma_{21} - k\times(q+\eta) \sigma_{31}] } } 
\e{ -\frac{i}{2} (\eta\times q + k\times \eta)} 
{\sst \frac{ (k-\eta)^{\rh2} \eta^{\rh3} q^{\rh4}
  (q+\eta)^{\rh5} }{(k-\eta)^2 \eta^2 q^2 (q+\eta)^2} }
\non \\
& & \si{\eta}{q}\si{k}{\eta}~ \epl{\m{31}\rh{2}\rh{3}}
\epl{\m{21}\rh{4}\rh{5}} \epl{\mu\nu\m{11}} 
\non \\
&=& - [3.01].
\eea

We note that this cancellation is essentially the same as one of those
presented in \cite{Kaminsky:2003}, because only one of the two
possible terms survives in the contraction of the two sources on
$W(k)$ into the {\it same} vertex:
\be
\epl{\m{1} \rh{A} \bl{2}} \epl{\m{2} \rh{B} \bl{3}} \epu{\bl1 \bl2
\bl3} \kt{1} \kt{2} = \delta^{\bl1}_{\m{2}} \epl{\m{1}\rh{A}\rh{B}}
\kt{1}\kt{2}.
\label{reduction}
\ee
This mechanism is completely general, and therefore by
induction, we can reduce any such pairs of graphs with $n+1$ sources on
$W(k)$, and zero sources from the path-ordered exponential
part of $O_{\mu\nu}$, to
the calculation of the $n$ source case, where $n\ge 2$.  
(For $n=1$, there is no pairing,
and we merely obtain the nonvanishing tree-level result.)  We will now
see how this same mechanism applies to a particular subset of
graphs with two sources on $W(k)$ and with one source from the
path-ordered exponential part of $O_{\mu\nu}$, and thereby recover the
other part of the
$O(g^3)$ calculation in \cite{Kaminsky:2003}.  Unfortunately, the
story becomes more complicated after that because sources from $W(k)$
will attach to {\it distinct} vertices.

%%%%%  Section 4  %%%%%
\section{Contributions from $O(g^2)$ terms in $W(k)$}
\setcounter{equation}{0}

%%%%% Figure 2 %%%%%
\begin{figure}
\centerline{\epsfxsize=1.9in\epsfbox{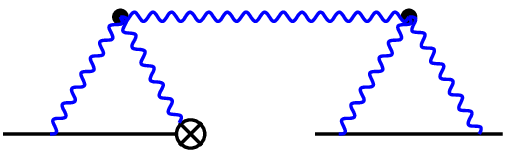} \hspace{0.1cm}
\epsfxsize=1.9in\epsfbox{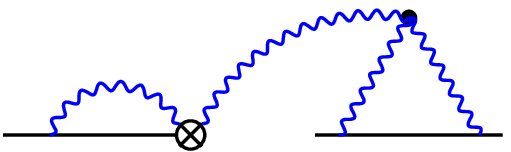} \hspace{0.1cm}
\epsfxsize=1.9in\epsfbox{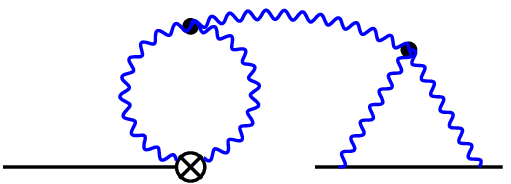} }
\caption{Diagrams [2.01], [2.02] and [2.03].  The first set of cancelling
contributions involving two gauge field sources from $W(k)$.}
\end{figure}

The two gauge field sources on $W(k)$ in the first three diagrams we
consider in this section connect to the same vertex as
shown in figure 2, and so we expect to find a cancellation amongst
them analogous to that in the previous section and the other part 
of calculation at $O(g^3)$ presented in \cite{Kaminsky:2003}.

Using (\ref{path simplification}) the first of these diagrams is given by
\bea
[2.01]&=& \frac{(ig)^3}{2}(-2ig)^2 \int \kt{11}\kt{21}
\tilde{k}^{\prime\mu_{12}}
\de{p_{11}+p_{21}-k} \de{p_{12}+p_{42}-k^{\prime}} \e{ -ip_{11}\cdot
\xi_{11}} \e{-\frac{i}{2} p_{11}\times p_{21}} \non \\
& &  \e{ -ip_{12}\cdot \xi_{12}} \e{-\frac{i}{2} p_{12}\times p_{42}}
\si{q_1}{q_2} \si{r_1}{r_2} {\sst \frac{ q_3^{\rh1} p_{11}^{\rh2} 
p_{21}^{\rh3} p_{12}^{\rh4} p_{42}^{\rh6} }{q_3^2 p_{11}^2 p_{21}^2
p_{12}^2 p_{42}^2} } \epu{\al1 \al2 \al3} \epu{\bl1 \bl2 \bl3}
(-ip_{42})_{ {\sst [} \mu} \epl{\nu {\sst ]} \rh6 \bl3}
\non \\
& &
\epl{\m{11}\rh2 \al1} \epl{\m{21}\rh3 \al2} \epl{\al3 \rh1 \bl1}
\epl{\m{12}\rh4 \bl2} 
\de{p_{11}+q_1} \de{p_{21}+q_2} \de{q_3+r_1} \de{p_{12}+r_2}
\de{p_{42}+r_3}.
\non \\ & & 
\eea
Now take $p_{11}= k-\eta$, $p_{21}= \eta$, $q_3=k$, $p_{12}=q$, 
$p_{42}= -(k+q)$, and contract on $\al2, \bl2$:
\bea
[2.01]&=&  -2 g^5 \int \kt{11}\kt{21}\kt{12} \e{i (k\times\eta)
(\sigma_{11}-\frac{1}{2}) } \e{i (k\times q)
(\sigma_{12}-\frac{1}{2}) } \si{k}{\eta} \si{k}{q}
{\sst \frac{ k^{\rh1} (k-\eta)^{\rh2} \eta^{\rh3}
q^{\rh4} (k+q)^{\rh6}}{k^2 (k-\eta)^2 \eta^2 q^2 (k+q)^2} } 
\non \\ & &
~ \times \epl{\m{11}\rh{2}\rh{3}} \epl{\m{21}\rh{1}\rh{4}} (k+q)_{[\mu}
\epl{\nu] \rh{6}\m{12}}
\non \\
&=& -2 g^5 \int \kt{11}\kt{21} \e{i (k\times\eta)
(\sigma_{11}-\frac{1}{2}) } \e{i (k\times q)
(\sigma_{12}-\frac{1}{2}) } \si{k}{\eta} \si{k}{q}
{\sst \frac{ k^{\rh1} (k-\eta)^{\rh2} \eta^{\rh3}
q^{\rh4} }{k^2 (k-\eta)^2 \eta^2 q^2 (k+q)^2} } 
\non \\ & &
~ \times
\epl{\m{11}\rh{2}\rh{3}} \epl{\m{21}\rh{1}\rh{4}} \left[
(k\times q)\epl{\mu\nu\rh{6}}(k+q)^{\rh6} - \epl{\mu\nu\m{12}}
\kt{12} (k+q)^2 \right],
\eea
where we have used $\tilde{k}^{\prime} = - \tilde{k}$, and again applied
(\ref{master identity}).  Note this time however, the term on the
right hand side  of (\ref{master identity}) survives because the
propagator associated with $p_{42}$ carries a loop momentum.  We will
need $[2.03]$ to cancel the additional term.

Like $[3.02]$, there are two contractions to consider in $[2.02]$:
\bea
[2.02]&=& -\frac{(ig)^4}{2}(-2ig) \int \kt{11} \kt{21}
\tilde{k}^{\prime\mu_{12}} \de{p_{11}+p_{21}-k}
\de{p_{12}+p_{32}+p_{42}-k^{\prime}} 
\e{-i p_{11}\cdot \xi_{11}} \e{-\frac{i}{2}p_{11}\times p_{21}}
\non \\ & &
\e{-i p_{12}\cdot \xi_{12}}
\e{-\frac{i}{2}[ p_{32}\times (p_{42}-k^{\prime})
- p_{42}\times k^{\prime}]} \si{q_1}{q_2} \epu{\al1 \al2 \al3}
\frac{ p_{11}^{\rh{2}} p_{21}^{\rh{3}} p_{12}^{\rh{4}} }{p_{11}^2
p_{21}^2 p_{12}^2} \bigg\{ \frac{p_{42}^{\rh{1}}}{p_{42}^2} 
\epl{\m{11}\rh{2}\al{1}} \epl{\m{21}\rh{3}\al{2}}
\non \\ & &
~\times \epl{\m{12}\rh{4}\mu} \epl{\nu\rh{1}\al{3}}
\de{p_{11}+q_1}
\de{p_{21}+q_2} \de{p_{12}+p_{32}} \de{p_{42}+q_3}
+ \frac{p_{32}^{\rh{1}}}{p_{32}^2} 
\epl{\m{11}\rh{2}\al{1}} \epl{\m{21}\rh{3}\al{2}}
\non \\ & & 
~\times \epl{\m{12}\rh{4}\nu} \epl{\mu\rh{1}\al{3}}
\de{p_{11}+q_1}
\de{p_{21}+q_2} \de{p_{12}+p_{42}} \de{p_{32}+q_3} \bigg\} - ( \mu
\leftrightarrow \nu).
\non \\ &&
\eea
As in $[2.01]$, set $p_{11}=k-\eta$, $p_{21}=\eta$, and $p_{12}=q$.
Then contracting on $\al2$ and using (\ref{epsilon identity 3}),
the remaining momentum constraints imply 
\bea
[2.02]&=& ig^5 \int \kt{11}\kt{21}\kt{12} \e{i (k\times \eta) 
(\sigma_{11}-\frac{1}{2})} \e{i(k\times q)\sigma_{12}} \si{k}{\eta}
{\sst \frac{ k^{\rh1} (k-\eta)^{\rh2} \eta^{\rh3} q^{\rh4}}{ k^2
(k-\eta)^2 \eta^2 q^2}}
\non \\ & &
\left\{ \epl{\m{11}\rh{2}\rh{3}} \epl{\m{12}\rh{4}\mu} \epl{\nu \rh{1}
\m{21}} + \epl{\m{11}\rh{2}\rh{3}} \epl{\m{12}\rh{4}\nu} \epl{\mu
\rh{1} \m{21}} \e{-i k\times q} \right\} - (\mu\leftrightarrow \nu)
\non \\ &=&
-2 g \int \kt{11}\kt{21}\kt{12} \e{i (k\times \eta) 
(\sigma_{11}-\frac{1}{2})} \e{i(k\times q)(\sigma_{12}-\frac{1}{2})} 
\si{k}{\eta} \si{k}{q}
{\sst \frac{ k^{\rh1} (k-\eta)^{\rh2} \eta^{\rh3} q^{\rh4}}{ k^2
(k-\eta)^2 \eta^2 q^2}} 
\non \\ & &
\epl{\m{11}\rh{2}\rh{3}}\epl{\m{21}\rh{1}\rh{4}} \epl{\mu\nu\m{12}}.
\eea

Now consider $[2.03]$:
\bea
[2.03]&=& -\frac{(ig)^3}{2}(-2ig)^2 \int \kt{11}\kt{21}
\de{p_{11}+p_{21}-k} \de{p_{32}+p_{42}-k^{\prime}} 
\e{-i p_{11}\cdot \xi_{11}} \e{-\frac{i}{2} p_{11}\times p_{21}}
\non \\ & &
\e{-\frac{i}{2} p_{32}\times p_{42}} \si{q_1}{q_2} \si{r_1}{r_2}
\epu{\al1 \al2 \al3}\epu{\bl1 \bl2 \bl3} {\sst \frac{p_{11}^{\rh2}
p_{21}^{\rh3} q_3^{\rh{1}} p_{32}^{\rh4} p_{42}^{\rh{6}}} {p_{11}^2
p_{21}^2 q_3^2 p_{32}^2 p_{42}^2} } 
\epl{\m{11}\rh{2}\al{1}}\epl{\m{21}\rh{3}\al{2}}
\non \\ & &
\epl{\al{3}\rh{1}\bl{1}}\epl{\mu\rh{4}\bl{2}} \epl{\nu\rh{6}\bl{3}} 
\de{p_{11}+q_1} \de{p_{21}+q_2} \de{q_3+r_1} \de{p_{32}+r_2}
\de{p_{42}+r_3} - {\sst (\mu\leftrightarrow\nu) }. 
\eea
Set $p_{11}=k-\eta$, $p_{21}=\eta$, $p_{32}=q$, $p_{42}=-(k+q)$, and
contract on $\al2, \bl2$:
\bea
[2.03]&=& 2ig^5 \int \kt{11}\kt{21} \e{i (k\times\eta)(\sigma_{11}-
\frac{1}{2})} \e{-\frac{i}{2}k\times q} \si{k}{\eta} \si{k}{q}
\frac{k^{\rh{1}} (k-\eta)^{\rh{2}} \eta^{\rh{3}} q^{\rh{4}}
(k+q)^{\rh{6}}} {k^2 (k-\eta)^2 \eta^2 q^2 (k+q)^2}
\non \\ & &
\epl{\m{11}\rh{2}\rh{3}} \left[ \epl{\m{21}\rh{1}\rh{4}}
\epl{\mu\nu\rh{6}} - \epl{\m{21}\rh{1}\mu}\epl{\nu\rh{6}\rh{4}}
\right] - (\mu\leftrightarrow\nu)
\non \\ &=&
 2ig^5 \int \kt{11}\kt{21} \e{i (k\times\eta)(\sigma_{11}-
\frac{1}{2})} \e{-\frac{i}{2}k\times q} \si{k}{\eta} \si{k}{q}
\frac{k^{\rh{1}} (k-\eta)^{\rh{2}} \eta^{\rh{3}} q^{\rh{4}}
(k+q)^{\rh{6}}} {k^2 (k-\eta)^2 \eta^2 q^2 (k+q)^2}
\non \\ & &
\epl{\m{11}\rh{2}\rh{3}} \left[ 2\epl{\m{21}\rh{1}\rh{4}}
\epl{\mu\nu\rh{6}} - ( -\epl{\m{21}\rh{1}\rh{6}} \epl{\mu\nu\rh{4}} +
 \epl{\m{21}\rh{1}\rh{4}}\epl{\mu\nu\rh{6}} ) \right]
\non \\ &=&
 2ig^5 \int \kt{11}\kt{21} \e{i (k\times\eta)(\sigma_{11}-
\frac{1}{2})} \e{-\frac{i}{2}k\times q} \si{k}{\eta} \si{k}{q}
\frac{k^{\rh{1}} (k-\eta)^{\rh{2}} \eta^{\rh{3}} q^{\rh{4}}
(k+q)^{\rh{6}}} {k^2 (k-\eta)^2 \eta^2 q^2 (k+q)^2}
\non \\ & &
\epl{\m{11}\rh{2}\rh{3}} \left[ \epl{\m{21}\rh{1}\rh{4}}
\epl{\mu\nu\rh{6}} +  \epl{\m{21}\rh{1}\rh{6}} \epl{\mu\nu\rh{4}}
\right]
\non \\ &=&
4 g^5 \int \kt{11}\kt{21} \e{i (k\times\eta)(\sigma_{11}-
\frac{1}{2})} \si{k}{\eta} \left[ \si{k}{q} \right]^2
\frac{k^{\rh{1}} (k-\eta)^{\rh{2}} \eta^{\rh{3}} q^{\rh{4}}
(k+q)^{\rh{6}}} {k^2 (k-\eta)^2 \eta^2 q^2 (k+q)^2}
\non \\ & &
\epl{\m{11}\rh{2}\rh{3}} \epl{\m{21}\rh{1}\rh{4}}
\epl{\mu\nu\rh{6}}.
\eea
In the second step we have used (\ref{epsilon identity 2}), and in the
fourth we have made the change of variables $q\rightarrow -(k+q)$ in the
second term.  This allows us to compare $[2.03]$ to the first term in
$[2.01]$, which is a total derivative with respect to the integration
over $\sigma_{12}$, whose measure we have suppressed.  Performing the 
integral, 
\be
\int_0^1 d\sigma_{12}~ (k\times q)~ \e{i (k\times
q)(\sigma_{12}-\frac{1}{2}) } = 2~ \sin{\left(\frac{k\times q}{2}\right)},
\non
\ee
we see that the first term in $[2.01]$ is cancelled by $[2.03]$, while the
second term in $[2.01]$ is cancelled by $[2.02]$ in the same way that $[3.02]$
cancels $[3.01]$.

Again, the identity (\ref{reduction}) ensures that this is essentially
the same calculation as that presented in \cite{Kaminsky:2003} at
one-loop.  However, the remaining graphs we consider in this section,
in which the two gauge field sources from $W(k)$ attach to different
vertices, work out quite differently.

%%%%% Figure 3 %%%%%
\begin{figure}
\centerline{\epsfxsize=2in\epsfbox{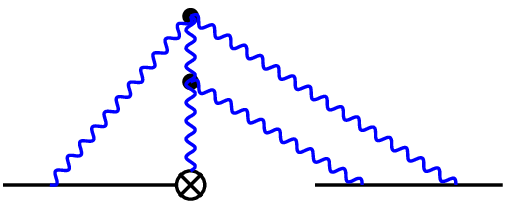} \hspace{1cm}
\epsfxsize=2in\epsfbox{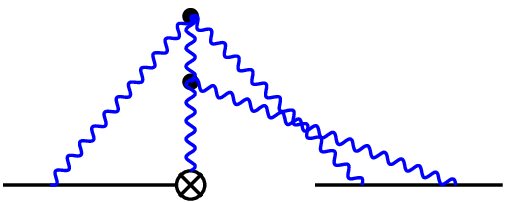}}
\caption{Diagrams [2.04a] and [2.04b]. The other two channels
corresponding to [2.01].}
\end{figure}

More specifically if we think of graph $[2.01]$ as containing a
tree-level four point function in the t-channel attached to the Wilson
lines to form a two-loop diagram, then the graphs in figure 3 reflect the
other two channels.  Combining the two we obtain
\bea
[2.04] &=& \frac{(ig)^3}{2} (-2ig)^2 \int \kt{11}\kt{21} 
\tilde{k}^{\prime\mu_{12}}
\de{p_{11}+p_{21}-k} \de{p_{12}+p_{42}-\kp} \e{-ip_{11}\cdot \xi_{11}}
\e{-\frac{i}{2} p_{11}\times p_{21}} \e{-ip_{12}\cdot \xi_{12}}
\non \\ & &
\times \e{-\frac{i}{2} p_{12}\times p_{42}}
\si{q_1}{q_2} \si{r_1}{r_2}
{\sst \frac{p_{11}^{\rh3} p_{21}^{\rh2} q_3^{\rh5}}
{p_{11}^2 p_{21}^2 q_3^2 } }
 \epu{\al1 \al2 \al3} \epu{\bl1 \bl2 \bl3} (-ip_{42})_{\mu}
\epl{\al3 \rh5 \bl1} \de{q_3+r_1}
\non \\ & &
\times
\epl{\m{11}\rh3 \al1} \epl{\m{21}\rh2 \bl2} \de{p_{11}+q_1}
\de{p_{21}+r_2} \left\{ {\sst \frac{p_{12}^{\rh4}p_{42}^{\rh6}}{p_{12}^2
p_{42}^2}} \epl{\m{12}\rh4 \al2} \epl{\nu\rh6 \bl3} \de{p_{12}+q_2}
\de{p_{42}+r_3} \right.
\non \\ & & ~
\left. +~ {\sst \frac{p_{12}^{\rh6} p_{42}^{\rh4}}{p_{12}^2
p_{42}^2}} \epl{\m{12}\rh6 \bl3} \epl{\nu\rh4 \al2} \de{p_{12}+r_3}
\de{p_{42}+q_2} \right\} - (\mu\leftrightarrow\nu).
\eea
Set $p_{11}=\eta$, $p_{12}=q$ in the first term and
$p_{12}=-(k+q)$ in the second, and contract on $\al1,
\bl1$ and $\al3, \bl3$ respectively, to get
\bea
[2.04]&=& -2g^5 \int \kt{11}\kt{21}\kt{12} \e{ik\times \eta
(\frac{1}{2}-\sigma_{11})} \si{\eta}{q} \bisi \proptwo
\non \\ & &
\left\{ {\sst \e{ik\times q (\sigma_{12}-\frac{1}{2})}} \epl{\m{21}\rh2 \rh5}
\epl{\m{12}\rh4 \rh3} (k+q)_{{\sst [}\mu} \epl{\nu {\sst ]} \rh6
\m{11}} - {\sst \e{ik\times q (\frac{1}{2}-\sigma_{12})} } 
\epl{\m{21}\rh2 \rh6}
\epl{\m{11}\rh3 \rh5} q_{{\sst [}\mu} \epl{\nu {\sst ]} \rh4
\m{12}} \right\} 
\non \\ &= & 2g^5 \int \kt{11}\kt{21}\kt{12} \e{ik\times \eta
(\frac{1}{2}-\sigma_{11})} \si{\eta}{q} \bisi \proptwo
\non \\ & &
\e{ik\times q (\frac{1}{2}-\sigma_{12})}
\left\{ \epl{\m{21}\rh2 \rh5}
\epl{\m{12}\rh3 \rh4} (k+q)_{{\sst [}\mu} \epl{\nu {\sst ]} \rh6
\m{11}} + \epl{\m{21}\rh2 \rh6}
\epl{\m{11}\rh3 \rh5} q_{{\sst [}\mu} \epl{\nu {\sst ]} \rh4
\m{12}} \right\}, 
\eea
where we have made the change of variables $\sigma_{12} \rightarrow
1-\sigma_{12}$ in the first term.  To these diagrams we pair up the 
ones in figure 4, which represent four sets of contractions:
%%%%% Figure 4 %%%%%
\begin{figure}
\centerline{\epsfxsize=2in\epsfbox{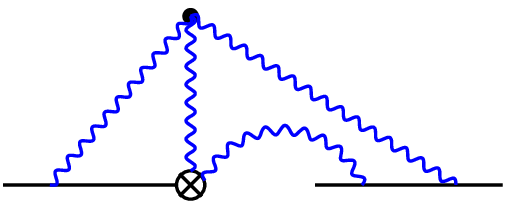} \hspace{1cm}
\epsfxsize=2in\epsfbox{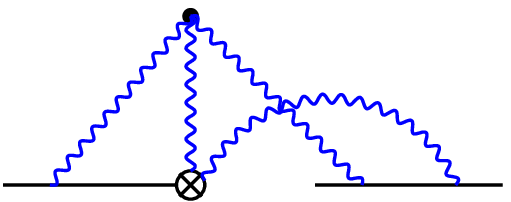}}
\caption{Diagrams [2.05a] and [2.05b].  Compare with [2.02].}
\end{figure}
%%%%%%%%%%%%%%%%%%%%
\bea
[2.05]&=& -\frac{(ig)^4}{2}(-2ig) \int \kt{11}\kt{21}
\tilde{k}^{\prime\mu_{12}} \de{p_{11}+p_{21}-k}
\de{p_{12}+p_{32}+p_{42}-\kp} \e{-i[p_{11}\cdot \xi_{11} +
\frac{1}{2}p_{11}\times p_{21}]}  
\non \\ & &
\e{-i p_{12}\cdot \xi_{12}}
\e{-\frac{i}{2}[ p_{32}\times (p_{42}-\kp) - p_{42}\times \kp]}
\si{q_1}{q_2} 
{\sst \frac{ p_{11}^{\rh3} p_{21}^{\rh2}}{p_{11}^2 p_{21}^2}} \bigg\{
{\sst \frac{p_{12}^{\rh{6}}}{p_{12}^2} } \epl{\m{21}\rh2 \al1} 
\epl{\m{12}\rh6 \al3} \de{p_{21}+q_1} \de{p_{12}+q_3}
\non \\ & &
\times \left[ {\sst \frac{p_{42}^{\rh5}}{p_{42}^2}} 
\epl{\m{11}\rh3 \mu} \epl{\nu \rh5 \al2} \de{p_{11}+p_{32}} \de{p_{42}+q_2}
+ {\sst \frac{p_{32}^{\rh5}}{p_{32}^2}} 
\epl{\m{11}\rh3 \nu} \epl{\mu \rh5 \al2} \de{p_{11}+p_{42}}
\de{p_{32}+q_2} \right] 
\non \\ & &
+~ {\sst \frac{p_{12}^{\rh{4}}}{p_{12}^2} } \epl{\m{11}\rh3 \al1} 
\epl{\m{12}\rh4 \al3} \de{p_{11}+q_1} \de{p_{12}+q_3} \left[ 
{\sst \frac{p_{42}^{\rh5}}{p_{42}^2}} 
\epl{\m{21}\rh2 \mu} \epl{\nu \rh5 \al2} \de{p_{21}+p_{32}}
\de{p_{42}+q_2}
\right.
\non \\ & & 
\left. + {\sst \frac{p_{32}^{\rh5}}{p_{32}^2}} 
\epl{\m{21}\rh2 \nu} \epl{\mu \rh5 \al2} \de{p_{21}+p_{42}}
\de{p_{32}+q_2}
\right] \bigg\} \epu{\al1 \al2 \al3} - (\mu\leftrightarrow\nu).
\eea
Take $p_{11}=\eta$.  In the first two terms set
$p_{12}=-(k+q)$, and in the second two set $p_{12}=q$.  Then
contracting on $\al3$ we obtain
\bea
[2.05]&=& -i g^5 \int \kt{11}\kt{21}\kt{12} \e{ik\times\eta(\frac{1}{2}
- \sigma_{11})} {\sst \frac{(k-\eta)^{\rh{2}} \eta^{\rh{3}}
(q+\eta)^{\rh{5}}} {(k-\eta)^2 \eta^2 (q+\eta)^2} } \epu{\al1 \al2
\al3} \bigg\{ \bisi {\sst \frac{(k+q)^{\rh{6}}}{(k+q)^2}}
\non \\ & &
\times~ 
\e{ik\times q(\frac{1}{2}- \sigma_{12})}
\epl{\m{21}\rh2 \al1} \epl{\m{12}\rh6 \al3}
\left[ \e{\frac{i}{2}\eta\times q} \epl{\m{11}\rh3 \mu} \epl{\nu
\rh5 \al2} + \e{-\frac{i}{2}\eta\times q} \epl{\m{11}\rh3 \nu} \epl{\mu
\rh5 \al2} \right] + 
\non \\ & &
~+ \si{\eta}{q} {\sst \frac{q^{\rh{4}}}{q^2}} 
\e{ik\times q(\sigma_{12}-\frac{1}{2})} 
\epl{\m{11}\rh3 \al1} \epl{\m{12}\rh4 \al3}
\left[ \e{\frac{i}{2} 
[\eta\times q - k\times q- k\times\eta] }
\epl{\m{21}\rh2 \mu} \epl{\nu \rh5 \al2} \right.
\non \\ & & 
\left. ~~ + \e{-\frac{i}{2}[ \eta\times
q-k\times q - k\times\eta]} \epl{\m{21}\rh2 \nu} \epl{\mu
\rh5 \al2} \right] \bigg\} - (\mu\leftrightarrow\nu)
\non \\ &=&
2 g^5 \int \kt{11}\kt{21}\kt{12} \e{ik\times\eta(\frac{1}{2}
- \sigma_{11})} \e{ik\times q(\frac{1}{2}- \sigma_{12})}
\bisi \si{\eta}{q} {\sst \frac{(k-\eta)^{\rh{2}} \eta^{\rh{3}}
(q+\eta)^{\rh{5}}} {(k-\eta)^2 \eta^2 (q+\eta)^2} }
\non \\ & &
\left\{ - {\sst \frac{(k+q)^{\rh6}}{(k+q)^2}} \epl{\m{21}\rh2 \rh6}
\epl{\m{11}\rh3 {\sst [}\mu} \epl{\nu {\sst ]} \rh5 \m{12}} 
-{\sst \frac{q^{\rh4}}{q^2}} \epl{\m{11}\rh3 \rh4} \epl{\m{21}\rh2
{\sst [}\mu} \epl{\nu {\sst ]} \rh5 \m{12}} \right\}
\non \\ &=&
2 g^5 \int \kt{11}\kt{21}\kt{12} \e{ik\times\eta(\frac{1}{2}
- \sigma_{11})} \e{ik\times q(\frac{1}{2}- \sigma_{12})}
\bisi \si{\eta}{q} {\sst \frac{(k-\eta)^{\rh{2}} \eta^{\rh{3}}
(q+\eta)^{\rh{5}}} {(k-\eta)^2 \eta^2 (q+\eta)^2} }
\non \\ & &
\left\{ {\sst \frac{q^{\rh4}}{q^2}} \epl{\m{11}\rh3 \rh4} \epl{\m{21}\rh2
\rh5} \epl{\mu\nu \m{12}} +
{\sst \frac{(k+q)^{\rh6}}{(k+q)^2}} \epl{\m{21}\rh2 \rh6}\epl{\m{11}\rh3
\rh5} \epl{\mu \nu \m{12}} \right\},
\eea
where we have used (\ref{epsilon identity 4}) in the last line.
Again, $[2.04]$ and $[2.05]$ do not cancel, but it is natural to combine
them using (\ref{master identity}) to obtain a total derivative term
with respect to the $\sigma_{12}$ integration:
\bea
[2.04] + [2.05] &=& 2g^5 \int \kt{11}\kt{21} \e{ik\times \eta
(\frac{1}{2}-\sigma_{11})} \si{\eta}{q} \bisi \proptwo
\non \\ & &
(k\times q) \e{ik\times q (\frac{1}{2}-\sigma_{12})} \left[
\epl{\m{11}\rh 3 \rh 4} \epl{\m{21}\rh2 \rh5}\epl{\mu\nu\rh6} +
\epl{\m{21}\rh 2 \rh 6} \epl{\m{11}\rh3 \rh5}\epl{\mu\nu\rh4} \right].
\label{204+205}
\eea

%%%%% Figure 5 %%%%%
\begin{figure}
\centerline{\epsfxsize=2in\epsfbox{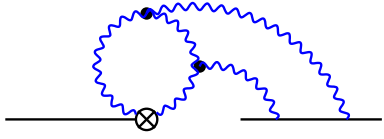}}
\caption{Diagram [2.06].}
\end{figure}
%%%%%%%%%%%%%%%%%%%%

In analogy to $[2.01]$-$[2.03]$ we now hope to cancel these terms with the
contribution in figure 5, which represents {\it two} sets of contractions (with
the same natural diagrammatic representation).  However, the fact that
the two sources from the commutator contract into distinct vertices will
seriously complicate the problem as we shall now see:
\bea
[2.06]&=& -\frac{(ig)^3}{2}(-2ig)^2 \int \kt{11}\kt{21}
\de{p_{11}+p_{21}-k} \de{p_{32}+p_{42}-\kp} \e{-ip_{11}\cdot \xi_{11}}
\e{-\frac{i}{2}p_{11}\times p_{21}} \e{-\frac{i}{2}p_{32}\times p_{42}}
\non \\ & &
{\sst \frac{p_{11}^{\rh3} p_{21}^{\rh2} q_3^{\rh5}}{p_{11}^2 p_{21}^2
q_3^2}} \si{q_1}{q_2}\si{r_1}{r_2} \epu{\al1 \al2 \al3} \epu{\bl1 \bl2
\bl3} \epl{\m{11}\rh3 \al1}\epl{\m{21}\rh2 \bl2} \epl{\al3 \rh5 \bl1}
\de{p_{11}+q_1}
\non \\ & &
\de{p_{21}+r_2}
\de{q_3+r_1} \left\{ {\sst \frac{p_{32}^{\rh4} p_{42}^{\rh6}} 
{p_{32}^2 p_{42}^2}}
\epl{\mu\rh4 \al2} \epl{\nu\rh6 \bl3} \de{p_{32}+q_2}
\de{p_{42}+r_3}  \right.
\non \\ & & \left. \qquad \qquad ~+~
{\sst \frac{p_{32}^{\rh6} p_{42}^{\rh4}}{p_{32}^2 p_{42}^2}}
\epl{\mu\rh6 \bl3}\epl{\nu\rh4 \al2} \de{p_{32}+r_3}
\de{p_{42}+q_2} \right\} 
 - {\sst (\mu\leftrightarrow\nu)}. 
\eea
Set $p_{11}=\eta$ and $p_{32} =q$ in the first
term, and $p_{32}=-(k+q)$ in the second.  Then contracting on
$\al1, \bl2$, we get
\bea
[2.06]&=& 2ig^5 \int \kt{11}\kt{21} \e{ik\times
\eta(\frac{1}{2}-\sigma_{11})} \proptwo \si{\eta}{q} \bisi
\epu{\al1 \al2 \al3}
\non \\ & &
\epu{\bl1 \bl2 \bl3} \epl{\al3 \rh5 \bl1}
\epl{\m{11}\rh3 \al1} \epl{\m{21}\rh2 \bl2}
\left\{ \epl{\mu\rh4 \al2}\epl{\nu\rh6 \bl3} \e{-\frac{i}{2}k\times q}
+ \epl{\mu\rh6 \bl3} \epl{\nu\rh4 \al2} \e{\frac{i}{2}k\times q}
\right\} - {\sst (\mu\leftrightarrow\nu) }
\non \\ &=&
- 4g^5 \int \kt{11}\kt{21} \e{ik\times \eta(\frac{1}{2}-\sigma_{11})}
\proptwo \si{\eta}{q} \si{k}{q} \bisi
\non \\ & &
\left[ \epl{\rh2 \rh3 \rh5} \epl{\mu\rh6 \m{21}} \epl{\nu\rh4 \m{11}} -
\epl{\m{21} \rh3 \rh5} \epl{\mu \rh6 \rh2}\epl{\nu\rh4 \m{11}} -
\epl{\m{11}\rh5 \rh2} \epl{\mu\rh6 \m{21}} \epl{\nu\rh4 \rh3} \right]
- {\sst (\mu\leftrightarrow\nu)}
\non \\ &=&
-4g^5 \int \kt{11}\kt{21} \e{ik\times \eta(\frac{1}{2}-\sigma_{11})}
\proptwo \si{\eta}{q} \si{k}{q} \bisi
\non \\ & &
~\times \left[ -\epl{\rh2 \rh3 \rh5} \epl{\m{11}\rh4 \rh6} \epl{\mu\nu\m{21}}
+ \epl{\m{21}\rh3 \rh5} \left( \epl{\m{11}\rh2 \rh6} \epl{\mu\nu\rh4}
+ \epl{\rh2 \rh4 \rh6} \epl{\mu\nu \m{11}} \right) \right. 
\non \\ & & \left. \qquad +~ 
\epl{\m{11}\rh2 \rh5} \left(
\epl{\m{21}\rh3 \rh4} \epl{\mu\nu\rh6} - \epl{\rh3 \rh4 \rh6}
\epl{\mu\nu\m{21}} \right) \right],
\eea
where we have used (\ref{epsilon identity 3}) and (\ref{epsilon
identity 4}).  Using the interchangeability of the $\m{11}$ and
$\m{21}$ indices, we now repeatedly apply (\ref{epsilon identity 2}) to
the first, third and fifth terms as follows:
\bea
& & [ -\epl{\rh2 \rh3 \rh5} \epl{\m{11} \rh4 \rh6} \epl{\mu\nu \m{21}} +
\epl{\m{21}\rh3 \rh5} \epl{\rh2 \rh4 \rh6} \epl{\mu\nu \m{11}} -
\epl{\rh3 \rh4 \rh6} \epl{\m{11}\rh2 \rh5} \epl{\mu\nu\m{21}} ]
\kt{11}\kt{21}
\non \\ &=&
[-\epl{\m{11}\rh4 \rh6} \epl{\rh2 \rh3 \rh5} + 
 \epl{\m{11}\rh2 \rh5} \epl{\rh3 \rh4 \rh6} + \epl{\m{11}\rh4 \rh5}
\epl{\rh3 \rh6 \rh2} + \epl{\m{11}\rh6 \rh5} \epl{\rh3 \rh2 \rh4}  
\non \\ & &
\qquad -\epl{\m{11} \rh2 \rh5} \epl{\rh3 \rh4 \rh6} ] \epl{\mu\nu\m{21}}
\kt{11} \kt{21}
\non \\ &=&
[ \epl{\rh2 \rh3 \rh5} \epl{\m{11}\rh6 \rh4} + 
\epl{\rh2 \rh3 \rh6} \epl{\m{11}\rh4 \rh5} + \epl{\rh2 \rh3 \rh4}
\epl{\m{11}\rh5 \rh6}  ]
\epl{\mu\nu\m{21}} \kt{11}\kt{21}
\non \\ &=&
\epl{\m{11}\rh2 \rh3} \epl{\rh4 \rh5 \rh6} \epl{\mu\nu\m{21}} \kt{11} 
\kt{21}.
\eea
Thus diagram $[2.06]$ is given by
\bea
[2.06]&=& -4g^5 \int \kt{11}\kt{21} \e{ik\times \eta(\frac{1}{2}-\sigma_{11})}
\proptwo \si{\eta}{q} \si{k}{q} \bisi
\non \\ & &
\times~ [ \epl{\m{21}\rh3 \rh5} \epl{\m{11}\rh2 \rh6} \epl{\mu\nu\rh4}
+ \epl{\m{11}\rh2 \rh5} \epl{\m{21}\rh3 \rh4} \epl{\mu\nu\rh6}
+ \epl{\m{11}\rh2 \rh3} \epl{\rh4 \rh5 \rh6} \epl{\mu\nu\m{21}} ],
\eea
and so evaluating the surface terms in (\ref{204+205}) with respect to
the $\sigma_{12}$ integration, and adding $[2.06]$, we obtain
\bea
[2.04]+[2.05]+[2.06]&=& -4g^5 \int \kt{11}\kt{21} 
\e{ik\times \eta(\frac{1}{2}-\sigma_{11})}
\proptwo \si{\eta}{q} \si{k}{q}
\non \\ & &
 \times~ \bisi \epl{\m{11}\rh2 \rh3} \epl{\rh4 \rh5 \rh6} 
\epl{\mu\nu\m{21}}.
\eea
This does not obviously vanish, and it is not immediately clear what can
be used to cancel this residual piece.  We will now show that the 
contributions in diagram 6, which contain one-loop vertex
corrections, precisely cancel this piece.

%%%%%  Figure 6  %%%%%
\begin{figure}
\centerline{\epsfxsize=2in\epsfbox{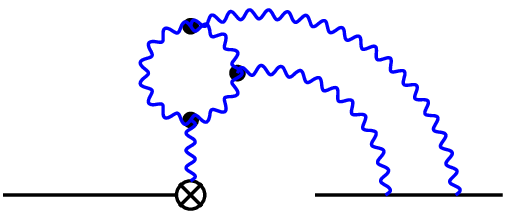} \hspace{1cm}
\epsfxsize=2in\epsfbox{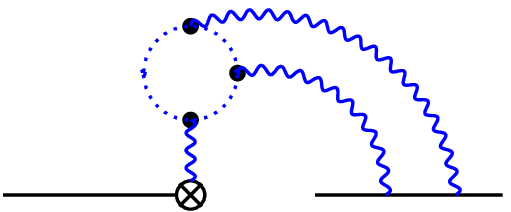}}
\caption{Diagrams [2.07] and [2.08], containing vertex corrections.}
\end{figure} 

Evaluating diagram $[2.07]$, we obtain
\bea
[2.07]&=& \frac{(ig)^2}{2}(-2ig)^3 \int \kt{11}\kt{21}
\de{p_{11}+p_{21}-k} \de{p_{42}-\kp} \e{-ip_{11}\cdot
\xi_{11}} \e{-\frac{i}{2}p_{12}\times p_{22}}
\si{q_1}{q_2} \si{r_1}{r_2}
\non \\ & &
\si{s_1}{s_2} \epu{\al1 \al2 \al3}
\epu{\bl1 \bl2 \bl3} \epu{\gl1 \gl2 \gl3} {\sst \frac{q_3^{\rh5}
r_3^{\rh6} s_3^{\rh4}}{q_3^2 r_3^2 s_3^2} } \epl{\al3 \rh5 \bl1}
\epl{\bl3 \rh6 \gl1} \epl{\gl3 \rh4 \al1} \de{q_3+r_1} \de{r_3+s_1}
\de{s_3+q_1}
\non \\ & &
\times (-ip_{42})_{\mu} 
{\sst \frac{ p_{11}^{\rh3} p_{21}^{\rh2} p_{42}^{\rh1}}
{p_{11}^2 p_{21}^2 p_{42}^2} }~ \epl{\m{11}\rh3 \al2} \epl{\m{21}\rh2
\bl2} \epl{\nu \rh1 \gl2} \de{p_{11}+q_2} \de{p_{21}+r_2}
\de{p_{42}+s_2} - {\sst (\mu\leftrightarrow\nu)}.
\eea
Take $p_{11}=\eta$, and $s_3=q$.  Then contract on $\al2$, $\bl2$:
\bea
[2.07]&=& 4g^5 \int \kt{11}\kt{21} \e{ik\times\eta
(\frac{1}{2}-\sigma_{11})} {\sst \frac{ k^{\rh1} \eta^{\rh3}
(k-\eta)^{\rh2} q^{\rh4} (q+\eta)^{\rh5} (k+q)^{\rh6}}{ k^2 \eta^2
(k-\eta)^2 q^2 (q+\eta)^2 (k+q)^2}} \si{\eta}{q} \si{k}{q} \bisi
\non \\ & &
\times \left(\delta^{\al3 \al1}_{\m{11}\rh3} - \delta^{\al3 \al1}_{\rh3
\m{11}} \right) \left( \delta^{\bl3 \bl1}_{\m{21}\rh2} - 
\delta^{\bl3 \bl1}_{\rh2
\m{21}} \right) \epu{\gl1 \gl2 \gl3} \epl{\al3 \rh5 \bl1} \epl{\bl3
\rh6 \gl1} \epl{\gl3 \rh4 \al1} k_{ {\sst [} \mu} \epl{\nu {\sst ]}
\rh1 \gl2}
\non \\ &=& 
4g^5 \int \kt{11}\kt{21} \e{ik\times\eta
(\frac{1}{2}-\sigma_{11})} {\sst \frac{ k^{\rh1} \eta^{\rh3}
(k-\eta)^{\rh2} q^{\rh4} (q+\eta)^{\rh5} (k+q)^{\rh6}}{ k^2 \eta^2
(k-\eta)^2 q^2 (q+\eta)^2 (k+q)^2}} \si{\eta}{q} \si{k}{q} \bisi
\non \\ & &
\times \left[ \epl{\m{11} \rh5 \rh2} \left( \delta^{\gl2 \gl3}_{\m{21}\rh6 } -
\delta^{\gl2 \gl3}_{\rh6 \m{21}} \right) \epl{\gl3 \rh4 \rh3} 
- \epl{\rh3 \rh5 \rh2} \epl{\m{21}\rh6 \gl1}	 
\left( \delta^{\gl1 \gl2}_{\rh4 \m{11}} - \delta^{\gl1
\gl2}_{\m{11} \rh4} \right) \right.
\non \\ & & \left.
~+~ \epl{\rh3 \rh5 \m{21}} \epl{\rh2 \rh6 \gl1} 
\left( \delta^{\gl1 \gl2}_{\rh4 \m{11}} - \delta^{\gl1
\gl2}_{\m{11} \rh4} \right) \right] k_{{\sst [}\mu} \epl{\nu {\sst ]}
\rh1 \gl2}
\non \\ &=& 
4g^5 \int \kt{11}\kt{21} \e{ik\times\eta
(\frac{1}{2}-\sigma_{11})} {\sst \frac{ k^{\rh1} \eta^{\rh3}
(k-\eta)^{\rh2} q^{\rh4} (q+\eta)^{\rh5} (k+q)^{\rh6}}{ k^2 \eta^2
(k-\eta)^2 q^2 (q+\eta)^2 (k+q)^2}} \si{\eta}{q} \si{k}{q} \bisi
\non \\ & &
\times~ \left[ \epl{\m{21}\rh5 \rh2} \epl{\rh6 \rh4 \rh3} k_{{\sst [ }
\mu} \epl{\nu {\sst ]} \rh1 \m{21}} - \epl{\m{11}\rh5 \rh2}
\epl{\m{21}\rh4 \rh3} k_{{\sst [}\mu} \epl{\nu {\sst ]} \rh1 \rh6} -
\epl{\rh3 \rh5 \rh2} \epl{\m{21}\rh6 \rh4} k_{{\sst [}\mu} \epl{\nu
{\sst ]} \rh1 \m{11}} \right.
\non \\ & &
\left. \qquad + \epl{\m{21}\rh3 \rh5} \epl{\rh2 \rh6 \rh4} 
k_{{\sst [}\mu} \epl{\nu {\sst ]} \rh1 \m{11}} - \epl{\m{21}\rh3 \rh5}
\epl{\rh2 \rh6 \m{11}} k_{{\sst [}\mu} \epl{\nu {\sst ]} \rh1 \rh4}
\right]
\non \\ &=&
4g^5 \int \kt{11}\kt{21} \e{ik\times\eta
(\frac{1}{2}-\sigma_{11})} {\sst \frac{ \eta^{\rh3}
(k-\eta)^{\rh2} q^{\rh4} (q+\eta)^{\rh5} (k+q)^{\rh6}}{ k^2 \eta^2
(k-\eta)^2 q^2 (q+\eta)^2 (k+q)^2}} \si{\eta}{q} \si{k}{q} \bisi
\non \\ & &
\times \left[ -\epl{\mu\nu \m{21}} k^2 \epl{\rh3 \rh4 \rh6}
\epl{\m{11}\rh2 \rh5} - \epl{\m{11}\rh2 \rh5} \epl{\m{21}\rh3 \rh4}
k^{\rh1} k_{{\sst [}\mu} \epl{\nu {\sst ]} \rh1 \rh6} - \epl{\mu\nu \m{11}}
k^2 \epl{\rh2 \rh3 \rh5} \epl{\m{21}\rh4 \rh6} \right.
\non \\ & &
\left. \qquad + \epl{\mu\nu\m{11}} k^2 \epl{\m{21}\rh3 \rh5} \epl{\rh2
\rh4 \rh6} - \epl{\m{11}\rh2 \rh6} \epl{\m{21}\rh3 \rh5} k^{\rh1} 
k_{{\sst [}\mu} \epl{\nu {\sst ]} \rh1 \rh4} \right],
\label{207}
\eea
where we have used (\ref{master identity}) on
the terms containing $\epl{\nu\rh1 \m{i1}}$.
Collecting the terms proportional to $k^2$, and applying (\ref{epsilon
identity 2}) we have
\bea
& & \left[ -\epl{\mu\nu \m{21}} \epl{\rh3 \rh4 \rh6} \epl{\m{11}\rh2 \rh5}
- \epl{\mu\nu \m{11}} \epl{\rh2 \rh3 \rh5} \epl{\m{21}\rh4 \rh6}
+ \epl{\mu\nu\m{11}} \epl{\m{21}\rh3 \rh5} \epl{\rh2
\rh4 \rh6} \right] \kt{11}\kt{21}
\non \\ &=& -
\epl{\mu\nu\m{21}} \left[ \epl{\m{11}\rh2 \rh5} \epl{\rh3 \rh4 \rh6} +
\epl{\m{11}\rh4 \rh6} \epl{\rh2 \rh3 \rh5} - \epl{\m{11}\rh2 \rh5}
\epl{\rh3 \rh4 \rh6} - \epl{\m{11}\rh4 \rh5} \epl{\rh3 \rh6 \rh2}
\right. 
\non \\ & & \left. 
\qquad - \epl{\m{11}\rh6 \rh5} \epl{\rh3 \rh2 \rh4} \right] \kt{11}\kt{21}
\non \\ &=&  \epl{\mu\nu\m{21}} \left[ \epl{\rh2 \rh3
\rh5}\epl{\m{11} \rh6 \rh4} + \epl{\rh2 \rh3 \rh6} \epl{\m{11}\rh4
\rh5} + \epl{\rh2 \rh3 \rh4} \epl{\m{11}\rh5 \rh6} \right]
\kt{11}\kt{21}
\non \\ &=&
\epl{\mu\nu\m{21}} \epl{\m{11}\rh2 \rh3} \epl{\rh4 \rh5 \rh6} \kt{11}
\kt{21}.  
\eea
Inserting this back into (\ref{207}) we obtain finally:
\bea
[2.07]&=& 4g^5 \int \kt{11}\kt{21} \e{ik\times\eta
(\frac{1}{2}-\sigma_{11})} {\sst \frac{ (k-\eta)^{\rh2} \eta^{\rh3} 
q^{\rh4} (q+\eta)^{\rh5} (k+q)^{\rh6}}{ (k-\eta)^2 \eta^2 q^2
(q+\eta)^2 (k+q)^2}} \si{\eta}{q} \si{k}{q} \bisi
\non \\ & &
\left[ \epl{\mu\nu\m{21}} \epl{\m{11}\rh2 \rh3} \epl{\rh4 \rh5 \rh6}
 - \epl{\m{11}\rh2 \rh5} \epl{\m{21}\rh3 \rh4}
\frac{k^{\rh1}}{k^2} k_{{\sst [}\mu} \epl{\nu {\sst ]} \rh1 \rh6} - 
\epl{\m{11}\rh2 \rh6} \epl{\m{21}\rh3 \rh5} \frac{k^{\rh1}}{k^2} 
k_{{\sst [}\mu} \epl{\nu {\sst ]} \rh1 \rh4} \right].
\non \\
\label{[207]}
\eea

To this contribution, we must add the ghost loop diagram from figure
6.  Remembering that once we fix the contractions among the gauge
fields, there are {\it two} sets of contractions among the ghost fields
corresponding to the two directions of ghost number flow, we have:
\bea
[2.08] &=&
- \frac{(ig)^2}{2}(-2ig)^3 \int \kt{11}\kt{21} \de{p_{11}+p_{21}-k}
\de{p_{42}-\kp}  \e{-ip_{11}\cdot
\xi_{11}} \e{-\frac{i}{2}p_{11}\times p_{21}} \si{q_1}{q_2}
\si{r_1}{r_2}
\non \\ & & ~\times
\si{s_1}{s_2} {\sst (-iq_1)^{\al{}} (-ir_1)^{\bl{}} (-is_1)^{\gl{}}} {\sst
\frac{i^3}{q_3^2 r_3^2 s_3^2} \frac{p_{11}^{\rh3} p_{21}^{\rh2}
p_{42}^{\rh1}}{p_{11}^2 p_{21}^2 p_{42}^2} }  
\epl{\m{11}\rh3 \al{}} \epl{\m{21}\rh2 \bl{}} 
(-ip_{42})_{ {\sst [} \mu} \epl{\nu {\sst ]} \rh1 \gl{}}
\de{p_{11}+q_2}
\non \\ & &
~ \times \de{p_{21}+r_2} \de{p_{42}+s_2} \left\{
\de{q_3+r_1} \de{r_3+s_1} \de{s_3+q_1} + \de{q_3 +s_1} \de{r_3+q_1}
\de{s_3+r_1} \right\},
\non \\
\eea
where we have included a leading minus sign for the fermion loop.
As usual we take $p_{11}=\eta$, and now we take $s_3=q$ in the first
term, and $s_1=q$ in the second, from which we immediately get:
\bea
[2.08]&=& 4g^5 \int \kt{11}\kt{21} \e{ik\times \eta
(\frac{1}{2}-\sigma_{11})} {\sst \frac{ (k-\eta)^{\rh2} \eta^{\rh3}
q^{\rh4} (q+\eta)^{\rh5} (k+q)^{\rh6}}{k^2 (k-\eta)^2 \eta^2 q^2
(q+\eta)^2 (k+q)^2}}  \si{\eta}{q} \si{k}{q} \bisi
\non \\ & &
~\times~ \left[ \epl{\m{11}\rh3 \rh4} \epl{\m{21} \rh2 \rh5}
k^{\rh1} k_{ {\sst [} \mu} \epl{\nu {\sst ]} \rh1 \rh6} +
\epl{\m{11}\rh3 \rh5} \epl{\m{21}\rh2 \rh6} k^{\rh1} k_{{\sst [} \mu}
\epl{\nu {\sst ]} \rh1 \rh4} \right].
\eea

The ghost graph $[2.08]$ cancels the last two terms in
(\ref{[207]}), while the first term in (\ref{[207]}) is exactly the 
negative of the residual piece from
diagrams $[2.04]$-$[2.06]$.  In summary we have found the cancellation:
\be
[2.04]+[2.05]+[2.06]+[2.07]+[2.08] = 0.
\ee
Combining this with the cancellation amongst $[2.01]$, $[2.02]$ and
$[2.03]$, we conclude that the sum of contributions from $O(g^2)$
terms in $W(k)$ to the correlator $\vev{W(k)O_{\mu\nu}(\kp)}$ at
$O(g^5)$ vanish.

%%%%%  Section 5  %%%%%
\section{Contributions from $O(g)$ terms in $W(k)$}
\setcounter{equation}{0}

We now turn to the most difficult case, where we have up to four
sources on $O_{\mu\nu}(\kp)$.  All of the contributions in this section
have one gauge field source from $W(k)$, and it will require all of
the graphs to produce a final cancellation.  We will further absorb the
common factor $\delta^{(3)}(p_{11}-k)$ associated with this source, into the 
integration symbol.

%%%%% Figure 7 %%%%%
\begin{figure}
\centerline{\epsfxsize=2in\epsfbox{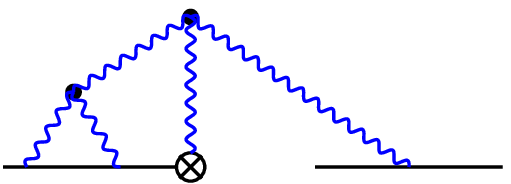} \hspace{1cm}
\epsfxsize=2in\epsfbox{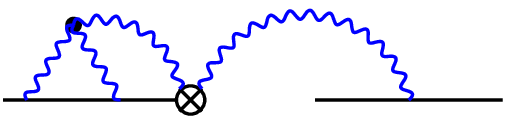} }
\caption{Diagrams [1.01] and [1.02].  Compare with [2.01] and [2.02].}
\end{figure}

First consider the contributions from figure 7.  The first is evaluated as
\bea
[1.01]&=& (ig)^3 (-2ig)^2 \int \kt{11}\kpt{12}\kpt{22}
\de{p_{12}+p_{22}+p_{42}-\kp} 
\e{-i[p_{12}\cdot \xi_{12} + p_{22}\cdot \xi_{22}]} \e{-\frac{i}{2}[
p_{22}\times (p_{42}-\kp) - p_{42}\times \kp]}
\non \\ & &
\si{q_1}{q_2} \si{r_1}{r_2} \epu{\al1 \al2 \al3} \epu{\bl1 \bl2 \bl3}
{\sst \frac{ p_{11}^{\rh1} p_{12}^{\rh3} p_{22}^{\rh5} p_{42}^{\rh6}
q_3^{\rh4}}{ p_{11}^2 p_{12}^2 p_{22}^2 p_{42}^2 q_3^2} }
(-ip_{42})_{\mu} \epl{\m{11}\rh1 \al1} \epl{ \nu
\rh6 \al2} \epl{\al3 \rh4 \bl1} \epl{\m{12} \rh3 \bl2}
\non \\ & &
\epl{\m{22}\rh5 \bl3} \de{p_{11}+q_1} \de{p_{42}+q_2} \de{q_3+r_1}
\de{p_{12}+r_2} \de{p_{22}+r_3} - {\sst (\mu\leftrightarrow\nu)}.
\eea
Setting $p_{12}= -\eta$, and $p_{22}= q+\eta$, and contracting on
$\al1, \bl3$ we obtain
\bea
[1.01]&=& 4g^5 \int \kt{11}\kt{12}\kt{22} \e{-i[k\times\eta \sigma_{12}
- k\times (q+\eta) \sigma_{22}]} \e{\frac{i}{2}[\eta\times q - k\times
q]} \si{k}{q}\si{\eta}{q} \propthree
\non \\ & &
\epl{\m{22}\rh1 \rh4} \epl{\m{12}\rh3 \rh5} (k+q)_{ {\sst
[} \mu} \epl{ \nu {\sst ]} \rh6 \m{11}}.
\eea
The second diagram in figure 7 represents two sets of 
contractions and is evaluated as
\bea
[1.02]&=& -(ig)^4 (-2ig) \int \kt{11} \kpt{12} \kpt{22}
\de{p_{12}+p_{22}+p_{32}+p_{42}-\kp} \e{-i[ p_{12}\cdot \xi_{12} +
p_{22} \cdot \xi_{22}]} \si{r_1}{r_2} \epu{\bl1 \bl2 \bl3}
\non \\ & &
\e{-\frac{i}{2}[ p_{22}\times (p_{32}+p_{42}-\kp) + p_{32}\times
(p_{42}-\kp) - p_{42}\times \kp]}
{\sst \frac{p_{11}^{\rh1} p_{12}^{\rh3} p_{22}^{\rh5}}{p_{11}^2
p_{12}^2 p_{22}^2}} \epl{\m{12}\rh3 \bl2} \epl{\m{22}\rh5 \bl3}
\de{p_{12}+r_2} \de{p_{22}+r_3}
\non \\ & &
\left\{ \epl{\m{11}\rh1 {\sst [} \mu} \epl{\nu {\sst ]}
\rh4 \bl1} \de{p_{11}+p_{32}} \de{p_{42}+r_1} {\sst
\frac{p_{42}^{\rh4}}{p_{42}^2}} +  \epl{\m{11}\rh1 {\sst [} \nu } 
\epl{\mu {\sst ]} \rh4 \bl1} \de{p_{11}+p_{42}} \de{p_{32}+r_1} {\sst
\frac{p_{32}^{\rh4}}{p_{32}^2}}  \right\}.
\non \\
\eea
Again take $p_{12}=-\eta$, $p_{22}=q+\eta$ and in the first term
take $p_{42}=-q$, while in the second take $p_{32}=-q$:
\bea
[1.02]&=& -2ig^5 \int \kt{11}\kt{12}\kt{22} \e{-i[k\times\eta
\sigma_{12} - k\times (q+\eta)\sigma_{22}]}
\e{\frac{i}{2}[\eta\times q - k\times q]} \si{\eta}{q}
{\sst \frac{ k^{\rh1} \eta^{\rh3} q^{\rh4} (q+\eta)^{\rh5}}{ k^2
\eta^2 q^2 (q+\eta)^2}}
\non \\ & &
\left[ \epl{\m{11}\rh1 \mu} \epl{\m{12}\rh3 \rh5} \epl{\nu\rh4 \m{22}}
\e{-\frac{i}{2} k\times q} + \epl{\m{11}\rh1 \nu} \epl{\m{12}\rh3
\rh5} \epl{\mu\rh4 \m{22}} \e{\frac{i}{2} k\times q} \right]
- {\sst (\mu\leftrightarrow\nu)}
\non \\ &=&
4 g^5 \int \kt{11}\kt{12}\kt{22} \e{-i[k\times\eta
\sigma_{12} - k\times (q+\eta)\sigma_{22}]}
\e{\frac{i}{2}[\eta\times q - k\times q]} \si{\eta}{q}
\si{k}{q}
\non \\ & &
{\sst \frac{ k^{\rh1} \eta^{\rh3} q^{\rh4} (q+\eta)^{\rh5}}{ k^2
\eta^2 q^2 (q+\eta)^2}}
\epl{\mu\nu\m{11}} \epl{\m{22}\rh1 \rh4} \epl{\m{12}\rh3 \rh5},
\eea
where we have used (\ref{epsilon identity 4}) in the last step.  
Combining $[1.01]$ and $[1.02]$ using (\ref{master identity}), we obtain
\bea
[1.01]+[1.02] &=& 4g^5 \int \kt{12}\kt{22}  \e{-i[k\times\eta
\sigma_{12} - k\times (q+\eta)\sigma_{22}]}
\e{\frac{i}{2}[\eta\times q - k\times q]} \si{\eta}{q}
\si{k}{q}
\non \\ & &
\propthree (k\times q) \epl{\m{22}\rh1 \rh4} \epl{\m{12}\rh3 \rh5}
\epl{\mu\nu\rh6}. 
\label{[101]+[102]}
\eea

%%%%% Figure 8 %%%%%
\begin{figure}
\centerline{\epsfxsize=2in\epsfbox{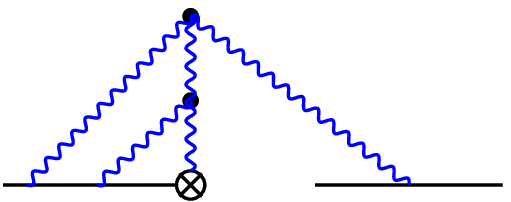} \hspace{1cm}
\epsfxsize=2in\epsfbox{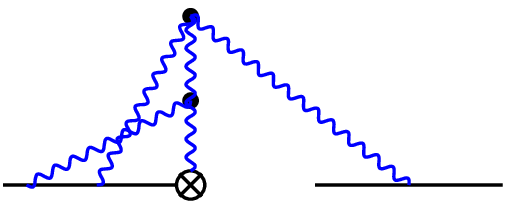} }
\caption{Diagrams [1.03a] and [1.03b].  Compare with [2.04a] and [2.04b].}
\end{figure}

Now consider the diagrams in figure 8, which, like the similar
diagrams in the previous section, correspond to the other
two channels with respect to $[1.01]$:
\bea
[1.03]&=& (ig)^3(-2ig)^2 \int \kt{11}\kpt{12}\kpt{22} 
\de{p_{12}+p_{22}+p_{42}-\kp} 
\e{-i[p_{12}\cdot \xi_{12} + p_{22}\cdot \xi_{22}]} \e{-\frac{i}{2}[
p_{22}\times (p_{42}-\kp) - p_{42}\times \kp]}
\non \\ & &
\si{q_1}{q_2} \si{r_1}{r_2} \epu{\al1 \al2 \al3} \epu{\bl1 \bl2 \bl3}
{\sst \frac{p_{11}^{\rh1} p_{42}^{\rh5} q_3^{\rh4} }{p_{11}^2 p_{42}^2
q_3^2} }
(-ip_{42})_{\mu} \epl{\m{11}\rh1 \al1} \epl{\al3 \rh4 \bl1}
\de{p_{11}+q_1} \de{q_3+r_1}
\non \\ & &
~\times~
\epl{\nu\rh5 \bl3} \de{p_{42}+r_3}
\left\{ \epl{\m{12}\rh6 \al2} \epl{\m{22}\rh3 \bl2} \de{p_{12}+q_2}
\de{p_{22}+r_2} {\sst \frac{ p_{12}^{\rh6} p_{22}^{\rh3}}{p_{12}^2
p_{22}^2}} \right.
\non \\ & & \hspace{4cm}
 \left.  + \epl{\m{12}\rh3 \bl2} \epl{\m{22}\rh6 \al2} 
\de{p_{12}+r_2} \de{p_{22}+q_2} 
{\sst \frac{ p_{12}^{\rh3} p_{22}^{\rh6}}{p_{12}^2
p_{22}^2}}  \right\} - {\sst (\mu\leftrightarrow\nu)}.
\eea
Set $p_{42}=q+\eta$, and $p_{22}= -\eta$ in the first term, 
$p_{12}= -\eta$ in the second.  Contracting on $\al1, \bl1$ yields
\bea
[1.03]&=& 4 g^5 \int \kt{11}\kt{12}\kt{22} \si{k}{q} \si{\eta}{q}
\propthree \epl{\m{12}\rh1 \rh6} \epl{\m{22}\rh3 \rh4} (q+\eta)_{ {\sst [} \mu}
\epl{\nu {\sst ]} \rh5 \m{11}}
\non \\ & &
\times \left\{ \e{-i[k\times q \sigma_{12} + k\times\eta
\sigma_{22}]} \e{\frac{i}{2}[\eta\times q + k\times q]} +
\e{-i[k\times\eta\sigma_{12} + k\times q \sigma_{22}]}
\e{\frac{i}{2} [ -\eta\times q + k\times q + 2k\times \eta]} \right\}. 
\eea

%%%%% Figure 9 %%%%%
\begin{figure}
\centerline{\epsfxsize=2in\epsfbox{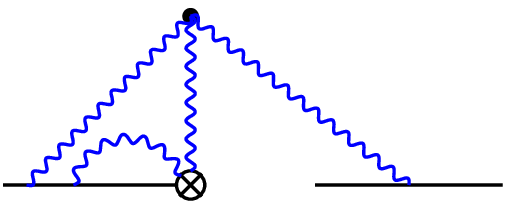} \hspace{1cm}
\epsfxsize=2in\epsfbox{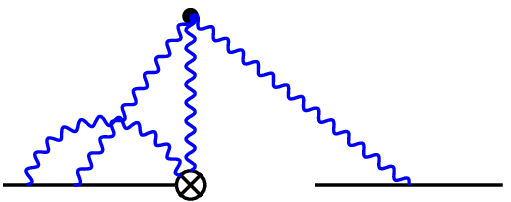} }
\caption{Diagrams [1.04a] and [1.04b].  Compare with [2.05a] and [2.05b].}
\end{figure}

As usual, we pair these together with the diagrams in figure 9 which
we denote by $[1.04a]$ and $[1.04b]$.  Writing out the four sets of 
contractions they represent, we get
\bea
[1.04]&=& -(ig)^4 (-2ig) \int \kt{11} \kpt{12} \kpt{22}
\de{p_{12}+p_{22}+p_{32}+p_{42}-\kp} \e{-i[ p_{12}\cdot \xi_{12} +
p_{22} \cdot \xi_{22}]} \si{r_1}{r_2} \epu{\bl1 \bl2 \bl3}
\non \\ & &
\e{-\frac{i}{2}[ p_{22}\times (p_{32}+p_{42}-\kp) + p_{32}\times
(p_{42}-\kp) - p_{42}\times \kp]} \si{r_1}{r_2} \epu{\bl1 \bl2 \bl3}
{\sst \frac{p_{11}^{\rh1}}{p_{11}^2}} \epl{\m{11}\rh1 \bl1}
\de{p_{11}+r_1} 
\non \\ & &
~\times \left\{ {\sst \frac{p_{12}^{\rh3} p_{22}^{\rh6}}{p_{12}^2 p_{22}^2}} 
\epl{\m{22}\rh6 \bl2}
\de{p_{22}+r_2} \bigg[ {\sst \frac{p_{42}^{\rh4}}{p_{42}^2}}
\epl{\m{12}\rh3 \mu}\epl{\nu\rh4 \bl3} \de{p_{12}+p_{32}}
\de{p_{42}+r_3} + {\sst \frac{p_{32}^{\rh4}}{p_{32}^2}}
\epl{\m{12}\rh3 \nu}\epl{\mu\rh4 \bl3} 
\right.
\non \\ & & 
~~~ \times 
\de{p_{12}+p_{42}} \de{p_{32}+r_3} 
\bigg] + {\sst \frac{p_{12}^{\rh6} p_{22}^{\rh3}}{p_{12}^2 p_{22}^2}} 
\epl{\m{12}\rh6 \bl2}
\de{p_{12}+r_2} \bigg[ {\sst \frac{p_{42}^{\rh4}}{p_{42}^2}}
\epl{\m{22}\rh3 \mu}\epl{\nu\rh4 \bl3} \de{p_{22}+p_{32}}
\non \\ & &
~~~ \left. \times \de{p_{42}+r_3} + {\sst \frac{p_{32}^{\rh4}}{p_{32}^2}}
\epl{\m{22}\rh3 \nu}\epl{\mu\rh4 \bl3} \de{p_{22}+p_{42}}
\de{p_{32}+r_3} \bigg] \right\} - {\sst (\mu\leftrightarrow\nu)}. 
\eea
Set $p_{12}=-\eta$, $p_{22}=-(k+q)$ in the first two terms and
$p_{12}=-(k+q)$, $p_{22}=-\eta$ in the second two.  Then
contract on $\bl1$, and use the symmetry under $\m{12}\leftrightarrow
\m{22}$ to obtain:
\bea
[1.04]&=& -2ig^5 \int \kt{11}\kt{12}\kt{22} \si{k}{q} {\sst \frac{
k^{\rh1} \eta^{\rh3} q^{\rh4} (k+q)^{\rh6}}{k^2 \eta^2 q^2
(k+q)^2}} \bigg[ \epl{\m{22}\rh1 \rh6} 
\e{-i [k\times\eta\sigma_{12} + k\times q \sigma_{22}]} 
\non \\ & &
\times \left( \e{\frac{i}{2}[-2\eta\times q + 2k\times \eta + k\times
q]} \epl{\m{12}\rh3 {\sst [}\mu} \epl{\nu {\sst ]}\rh4 \m{11}} +
\e{\frac{i}{2}[ 2k\times\eta + k\times q]} \epl{\m{12}\rh3 {\sst
[}\nu} \epl{\mu {\sst ]} \rh4 \m{11}} \right) + \epl{\m{12}\rh1 \rh6} 
\times
\non \\ & & 
\times \e{-i[ k\times q \sigma_{12} + k\times\eta\sigma_{22}]}
\left( \e{\frac{i}{2}k\times q} \epl{\m{22}\rh3 {\sst [} \mu} \epl{\nu
{\sst ]} \rh4 \m{11}} + \e{\frac{i}{2}[k\times q + 2\eta\times q]}
\epl{\m{22}\rh3 {\sst [}\nu} \epl{\mu {\sst ]} \rh4 \m{11}} \right) \bigg]
\non \\ &=&
4g^5 \int \kt{11}\kt{12}\kt{22} \si{k}{q} \si{\eta}{q} {\sst \frac{
k^{\rh1} \eta^{\rh3} q^{\rh4} (k+q)^{\rh6}}{k^2 \eta^2 q^2
(k+q)^2}} \epl{\m{12}\rh1 \rh6} \epl{\m{22}\rh3 \rh4}
\epl{\mu\nu\m{11}}
\non \\ & &
\times
\left\{ \e{-i[k\times\eta\sigma_{12}+k\times q \sigma_{22}]}
\e{\frac{i}{2}[-\eta\times q  +k\times q +2k\times \eta]} +
\e{-i[k\times q\sigma_{12}+k\times\eta\sigma_{22}]}
\e{\frac{i}{2}[\eta\times q + k\times q]} \right\},
\eea
where we have used (\ref{epsilon identity 4}) in the final step.

Summing the contributions in figures 8 and 9, and invoking
(\ref{master identity}) as usual we obtain
\bea
[1.03]+[1.04]&=& 4g^5 \int \kt{12}\kt{22} \si{k}{q} \si{\eta}{q}
\propthree {\sst k\times (q+\eta)} 
\epl{\m{12}\rh1 \rh6} \epl{\m{22}\rh3 \rh4} \epl{\mu\nu \rh5}
\non \\ & &
\times \left\{ \e{-i[k\times q \sigma_{12} + k\times \eta
\sigma_{22}]} \e{\frac{i}{2}[\eta\times q + k\times q]} +
\e{-i[k\times\eta\sigma_{12} + k\times q \sigma_{22}]}
\e{\frac{i}{2} [ -\eta\times q + k\times q + 2k\times \eta]} \right\}. 
\non \\
\label{[103]+[104]}
\eea

%%%%% Figure 10 %%%%%
\begin{figure}
\centerline{\epsfxsize=2in\epsfbox{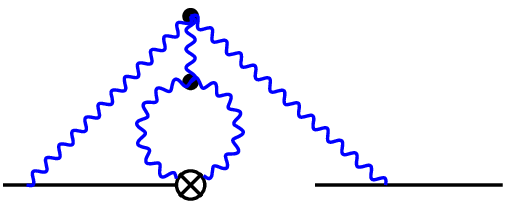} \hspace{1cm}
\epsfxsize=2in\epsfbox{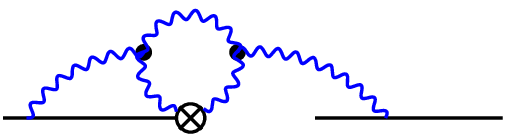} }
\caption{Diagrams [1.05] and [1.06].  Compare with [2.03] and [2.06].}
\end{figure}

Analogous to $[2.03]$ and $[2.06]$, we now have the diagrams given 
in figure 10, and which we denote  by $[1.05]$
and $[1.06]$ respectively.  However, in the last section $[2.03]$
precisely cancelled $[2.01]$ and $[2.02]$, in a calculation very similar
to that presented in \cite{Kaminsky:2003}.  On the other hand we will
find that $[1.05]$ by itself does {\it not} cancel the equivalent graphs
here, $[1.01]$ and $[1.02]$, but serves to 'correct' a noncommutative
phase in other diagrams.

Evaluating $[1.05]$, we find that
\bea
[1.05]&=& -(ig)^3 (-2ig)^2 \int \kt{11}\kpt{12}
\de{p_{12}+p_{32}+p_{42}-\kp} \e{-ip_{12}\cdot\xi_{12}}
\e{-\frac{i}{2} [p_{32}\times (p_{42}-\kp) - p_{42}\times \kp]}
\non \\ & &
\si{q_1}{q_2}\si{r_1}{r_2} \epu{\al1 \al2 \al3} \epu{\bl1 \bl2 \bl3}
{\sst \frac{p_{11}^{\rh1} p_{12}^{\rh6} p_{32}^{\rh5} p_{42}^{\rh3}
q_3^{\rh4}}{p_{11}^2 p_{12}^2 p_{32}^2 p_{42}^2 q_3^2}}
\epl{\m{11}\rh1 \al1} \epl{\m{12}\rh6 \al2} \epl{\mu\rh5 \bl2}
\epl{\nu\rh3 \bl3} \epl{\al3 \rh4 \bl1}
\non \\ & &
\de{p_{11}+q_1} \de{p_{12}+q_2} \de{p_{32}+r_2} \de{p_{42}+r_3}
\de{q_3+r_1} - (\mu\leftrightarrow\nu).
\eea
So as not to introduce a propagator carrying the momentum $k-\eta$,
set $p_{12}= -(k+q)$, $p_{42}=-\eta$, and contract on $\al1,\bl1$:
\bea
[1.05]&=& 4ig^5 \int \kt{11}\kt{12} \e{-ik\times q \sigma_{12}}
\e{-\frac{i}{2}(\eta\times q - k\times q)} \si{k}{q}\si{\eta}{q}
\propthree 
\non \\ & &
~\times \bigg\{ \epl{\m{12}\rh6 \rh1} \epl{\rh3 \rh4 {\sst [} \mu}
\epl{\nu {\sst ]} \rh5 \m{11}} + \epl{\m{12}\rh6 \rh1} \epl{\rh5 \rh4
{\sst [} \mu} \epl{\nu {\sst ]} \rh3 \m{11}} \bigg\}
\non \\ &=&
- 4ig^5 \int \kt{11}\kt{12} \e{-ik\times q \sigma_{12}}
\e{-\frac{i}{2}(\eta\times q - k\times q)} \si{k}{q}\si{\eta}{q}
\propthree 
\non \\ & &
~\times \epl{\m{11}\rh1 \rh6} \bigg\{ \epl{\m{12}\rh3 \rh4}
\epl{\mu\nu\rh5} + \epl{\m{12}\rh5 \rh4} \epl{\mu\nu\rh3} \bigg\}
\non \\ &=&
- 8g^5 \int \kt{11}\kt{12} \e{ik\times q(\frac{1}{2}- \sigma_{12})}
\si{k}{q} \left[\si{\eta}{q}\right]^2 \propthree 
\non \\ & &
~\times \epl{\m{11}\rh1 \rh6} \epl{\m{12}\rh3 \rh4}
\epl{\mu\nu\rh5}, 
\label{[105]}
\eea 
where we have invoked (\ref{epsilon identity 3}) twice to obtain the
second line and cancelled the resulting crossterm proportional to
$\epl{\mu\nu\m{12}}$, and where we have changed variables $\eta
\rightarrow -(q+\eta)$ in the second term on the second line, which effectively
interchanges the $\rh3$ and $\rh5$ indices.

Now consider $[1.06]$, which is similar to $[2.06]$ and represents two
sets of contractions with one natural diagrammatic representation.  Evaluating
it we obtain
\bea
[1.06]&=& -(ig)^3 (-2ig)^2 \int \kt{11}\kpt{12}
\de{p_{12}+p_{32}+p_{42}-\kp} \e{-ip_{12}\cdot
\xi_{12}} \e{-\frac{i}{2}[p_{32}\times (p_{42}-\kp)- p_{42}\times
\kp]} \si{q_1}{q_2} 
\non \\ & & 
\si{r_1}{r_2}   
{\sst \frac{ p_{11}^{\rh1} p_{12}^{\rh3} q_3^{\rh4}}{p_{11}^2 p_{12}^2
q_3^2}} \epu{\al1 \al2 \al3} \epu{\bl1 \bl2 \bl3} \epl{\al3 \rh4 \bl1}
\epl{\m{11}\rh1 \al1} \epl{\m{12}\rh3 \bl2} \de{q_3+r_1}
\de{p_{11}+q_1} \de{p_{12}+r_2} 
\non \\ & & 
 \left\{ {\sst \frac{p_{32}^{\rh6} p_{42}^{\rh5}}{p_{32}^2 p_{42}^2} }
\epl{\mu\rh6 \al2} \epl{\nu\rh5 \bl3} \de{p_{32}+q_2} \de{p_{42}+r_3}
+ {\sst \frac{p_{32}^{\rh5} p_{42}^{\rh6}}{p_{32}^2 p_{42}^2} }
\epl{\mu\rh5 \bl3} \epl{\nu\rh6 \al2} \de{p_{32}+r_3} \de{p_{42}+q_2}
\right\} 
\non \\ & & - (\mu\leftrightarrow\nu).
\eea
Set $p_{12}=-\eta$, and $p_{32}= -(k+q)$ in the first term and
$p_{42}=-(k+q)$ in the second.  Contracting on $\al1, \bl2$ we get
\bea
[1.06]&=& 4ig^5 \int \kt{11}\kt{12} \e{-ik\times \eta\sigma_{12}}
\si{\eta}{q} \si{k}{q} \propthree {\sst \left( \delta^{\al2
\al3}_{\m{11}\rh1} - \delta^{\al2 \al3}_{\rh1 \m{11}} \right)}
\non \\ & &
{\sst \left( \delta^{\bl3 \bl1}_{\m{12}\rh3} - \delta^{\bl3
\bl1}_{\rh3 \m{12}} \right) }
\epl{\al3 \rh4 \bl1} \left[ \epl{\al2 \rh6 [\mu} \epl{\nu] \rh5 \bl3}
\e{\frac{i}{2} [2k\times \eta + k\times q - \eta\times q]} +
\epl{\bl3\rh5 [\mu} \epl{\nu]\rh6 \al2} \e{\frac{i}{2}[ \eta\times q -
k\times q]} \right]
\non \\ &=& 
8 g^5 \int \kt{11}\kt{12} \e{ik\times \eta(\frac{1}{2}-\sigma_{12})}
\si{\eta}{q} \si{k}{q} \bisi \propthree 
\non \\ & &
\left[ \epl{\m{11}\rh6 [\mu}
\epl{\nu] \rh5 \m{12}} \epl{\rh1 \rh4 \rh3}
- \epl{\m{11}\rh6 [\mu} \epl{\nu] \rh5 \rh3} \epl{\rh1 \rh4 \m{12}} -
\epl{\rh1 \rh6 [\mu} \epl{\nu] \rh5 \m{12}} \epl{\m{11}\rh4 \rh3}
\right]
\non \\ &=&
8 g^5 \int \kt{11}\kt{12} \e{ik\times \eta(\frac{1}{2}-\sigma_{12})}
\si{\eta}{q} \si{k}{q} \bisi \propthree 
\non \\ & &
\left[ - \epl{\mu\nu\m{12}} \epl{\m{11}\rh6 \rh5} \epl{\rh1 \rh4 \rh3}
+ \epl{\mu\nu\rh6} \epl{\m{11}\rh3 \rh5} \epl{\m{12}\rh1 \rh4} -
\epl{\mu\nu\m{11}} \epl{\rh3 \rh5 \rh6} \epl{\m{12}\rh1 \rh4} \right.
\non \\ & & 
\left. ~~+ \epl{\mu\nu\rh5} \epl{\m{12}\rh1 \rh6} \epl{\m{11}\rh3
\rh4} - \epl{\mu\nu\m{12}} \epl{\m{11}\rh3 \rh4} \epl{\rh5 \rh1 \rh6}
\right]
\non \\ &=&
8 g^5 \int \kt{11}\kt{12} \e{ik\times \eta(\frac{1}{2}-\sigma_{12})}
\si{\eta}{q} \si{k}{q} \bisi \propthree 
\non \\ & &
\left[ \epl{\mu\nu\rh5} \epl{\m{12}\rh1 \rh6} \epl{\m{11}\rh3 \rh4} +
\epl{\mu\nu\rh6} \epl{\m{12}\rh1 \rh4} \epl{\m{11}\rh3 \rh5} -
\epl{\mu\nu\m{12}} \epl{\m{11}\rh1 \rh3} \epl{\rh4 \rh5 \rh6} \right].
\label{[106]}
\eea

Examining the graphs we have computed so far in (\ref{[101]+[102]}),
(\ref{[103]+[104]}), (\ref{[105]}) and (\ref{[106]}), we find ourselves
in a similar situation as that after having computed $[2.06]$.
However, while the tensor structures are now in the same form across
these contributions, the noncommutative phases are not manifestly the
same, as they were in the equivalent point in the previous section, and a
little more work is required.  Denote the sum of these graphs,
$[1.01]$-$[1.06]$, by $S$.  First examine terms that are proportional to 
$\epl{\mu\nu\rh6}$, which occur in (\ref{[101]+[102]}), and (\ref{[106]}):
\bea
[1.01]+[1.02] &= & 4g^5 \int \kt{12}\kt{22}  \e{-i[k\times\eta
\sigma_{12} - k\times (q+\eta)\sigma_{22}]}
\e{\frac{i}{2}[\eta\times q - k\times q]} \si{\eta}{q}
\si{k}{q}
\non \\ & &
\propthree (k\times q) \epl{\m{22}\rh1 \rh4} \epl{\m{12}\rh3 \rh5}
\epl{\mu\nu\rh6}, 
\non
\eea
\bea 
[1.06] & \supset & 8 g^5 
\int \kt{11}\kt{12} \e{ik\times \eta(\frac{1}{2}-\sigma_{12})}
\si{\eta}{q} \si{k}{q} \bisi
\non \\ & &
~\propthree \epl{\m{12}\rh1 \rh4} \epl{\m{11}\rh3 \rh5} \epl{\mu\nu\rh6}.
\eea
To proceed,
perform the change of variables $\eta\rightarrow -(q+\eta)$ in
each of these terms, and then average the original form and the new forms.
Then we may write
\bea
[1.01]+[1.02] &=& 2g^5 \int \kt{11}\kt{12} \si{\eta}{q}\si{k}{q}
\propthree \epl{\m{12}\rh1 \rh4} \epl{\m{11}\rh3 \rh5}
\epl{\mu\nu\rh 6}
\non \\ & & {\sst (k\times q)}
\left\{ \e{-i[k\times\eta\sigma_{12}-k\times (q+\eta)\sigma_{22}]}
\e{\frac{i}{2} [\eta\times q - k\times q]} + \e{i[k\times
(q+\eta)\sigma_{12} - k\times \eta\sigma_{22}]}
\e{-\frac{1}{2}[\eta\times q + k\times q]} \right\} ,
\non
\eea
while the term from $[1.06]$ can be written as
\bea
[1.06] &\supset& 4g^5 \int \kt{11}\kt{12}
\si{\eta}{q} \si{k}{q}
\propthree \epl{\m{12}\rh1 \rh4} \epl{\m{11}\rh3 \rh5}
\epl{\mu\nu\rh6}
\non \\ & &
\left\{ \e{ik\times\eta(\frac{1}{2}-\sigma_{12})} \bisi + \e{ik\times
(q+\eta)(\sigma_{12}-\frac{1}{2})} {\sst \sin{(\frac{-\eta\times q +
k\times \eta}{2})}} \right\}.
\eea

By performing the integrals over $p_{12}$ and $p_{22}$, and doing some
trigonometry, we can show that these are the
negatives of each other and so cancel.  Denoting $\eta\times q = A$,
$k\times q = B$ and $k\times \eta = C$ we obtain
\bea 
& &
2 k\times q \int
\left\{ \e{-i[k\times\eta\sigma_{12}-k\times (q+\eta)\sigma_{22}]}
\e{\frac{i}{2} [\eta\times q - k\times q]} + \e{i[k\times
(q+\eta)\sigma_{12} - k\times \eta\sigma_{22}]}
\e{-\frac{1}{2}[\eta\times q + k\times q]} \right\}
\non \\ &=&
\frac{2}{C (B+C)} \left\{ - C \e{\itwo
(A+B)} + C \e{\itwo (A-B )} - B \e{\itwo (A - B - 2 C)} + B \e{\itwo
(A-B)} \right.
\non \\ & & ~
\left.
+ (B+C) \e{-\itwo (A-B)} - (B+C)\e{-\itwo (A+B)} - B\e{-\itwo (A-B-2C)}
+ B \e{-\itwo (A+B)} \right\}
\non \\ &=&
\frac{8}{B+C} \sin{(\frac{A}{2})} \sin{(\frac{B}{2})} -
\frac{4B}{C(B+C)} \left[
\cos{(\frac{A-B-2C}{2})}  -  \cos{(\frac{A-B}{2})} \right],
\eea
while
\bea
& & 4 \int \e{ik\times\eta(\frac{1}{2}-\sigma_{12})} \bisi + \e{ik\times
(q+\eta)(\sigma_{12}-\frac{1}{2})} {\sst \sin{(\frac{-\eta\times q +
k\times \eta}{2})}} 
\non \\ &=&
\frac{8}{C} \sin{(\frac{C}{2})} \sin{(\frac{A-B-C}{2})} +
\frac{8}{B+C} \sin{(\frac{B+C}{2})} \sin{(\frac{C-A}{2})}
\non \\ &=&
\frac{4}{C} \left[ {\sst \cos{(\frac{A-B-2C}{2})}} - {\sst
\cos{(\frac{A-B}{2})}} \right] + \frac{4}{B+C} \left[ {\sst
\cos{(\frac{A+B}{2})}} - {\sst \cos{(\frac{-A+B+2C}{2})}} \right]
\non \\ &=&
\frac{4 B}{C(B+C)} \left[ \cos{(\frac{A-B-2C}{2})} -
\cos{(\frac{A-B}{2})} \right] - \frac{8}{B+C} \sin{(\frac{A}{2})}
\sin{(\frac{B}{2})}, 
\eea
using $2 \sin(x) \sin(y) = \cos(x-y)- \cos(x+y)$.  Thus the terms in
$S$ proportional to $\epl{\mu\nu\rh6}$ cancel.  Now consider the terms 
in $S$ proportional to $\epl{\mu\nu\rh5}$,
which occur in (\ref{[103]+[104]}), (\ref{[105]}), and (\ref{[106]}):
\bea
[1.03]+[1.04]&=& 4g^5 \int \kt{12}\kt{22} \si{k}{q} \si{\eta}{q}
\propthree {\sst k\times (q+\eta)} 
\epl{\m{12}\rh1 \rh6} \epl{\m{22}\rh3 \rh4} \epl{\mu\nu \rh5}
\non \\ & &
\times \left\{ \e{-i[k\times q \sigma_{12} + k\times \eta
\sigma_{22}]} \e{\frac{i}{2}[\eta\times q + k\times q]} +
\e{-i[k\times\eta\sigma_{12} + k\times q \sigma_{22}]}
\e{\frac{i}{2} [ -\eta\times q + k\times q + 2k\times \eta]} \right\}, 
\non
\eea
\bea
[1.05]&=& - 8g^5 \int \kt{11}\kt{12} \e{ik\times q(\frac{1}{2}- \sigma_{12})}
\si{k}{q} \left[\si{\eta}{q}\right]^2 \propthree 
\non \\ & &
~\times \epl{\m{11}\rh1 \rh6} \epl{\m{12}\rh3 \rh4}
\epl{\mu\nu\rh5},
\non
\eea
\bea
[1.06]&\supset& 8 g^5 \int \kt{11}\kt{12} \e{ik\times \eta(\frac{1}{2}-\sigma_{12})}
\si{\eta}{q} \si{k}{q} \bisi \propthree 
\non \\ & &
\epl{\m{11}\rh1 \rh6} \epl{\m{12}\rh3 \rh4} \epl{\mu\nu\rh5}.
\eea
We need not apply the trick that we applied for the $\epl{\mu\nu\rh6}$
terms here, and can directly evaluate the phases.  The phase from
$[1.03]+[1.04]$ is evaluated to be
\bea
& &
4 k\times (q+\eta) \int \e{-i[k\times q \sigma_{12} + k\times \eta
\sigma_{22}]} \e{\frac{i}{2}[\eta\times q + k\times q]} +
\e{-i[k\times\eta\sigma_{12} + k\times q \sigma_{22}]}
\e{\frac{i}{2} [ -\eta\times q + k\times q + 2k\times \eta]}
\non \\ &=&
\frac{4}{BC}\left[ B \e{\itwo (A+B)} - B \e{\itwo (A-B-2C)} + C
\e{\itwo (-A+B+2C)} - C \e{-\itwo (A+B)} \right.
\non \\ & & ~
\left. + (B+C) \left( \e{\itwo (A-B)} + \e{\itwo (B-A)}\right) - (B+C)
\e{\itwo (A+B)} - (B+C) \e{\itwo (-A+B+2C)} \right]
\non \\ &=& 
\frac{4}{BC} \left\{ 4C \sin{(\frac{A}{2})}\sin{(\frac{B}{2})} + 2B
\left[ \cos{(\frac{A-B}{2})} - \cos{(\frac{A-B-2C}{2})} \right] \right\}, 
\eea
while the phase from $[1.05]$ is given by
\be
- 8 \int \e{ik\times q(\frac{1}{2}-\sigma_{12})} \si{\eta}{q} = -\frac{16}{B}
\sin{(\frac{B}{2})} \sin{(\frac{A}{2})},
\ee
and the phase from $[1.06]$ is given by
\bea
8 \int \e{ik\times \eta (\frac{1}{2}-\sigma_{12})} \bisi &=& 
\frac{16}{C} \sin{(\frac{C}{2})} \sin{(\frac{A-B-C}{2})} \
\non \\ &=&
\frac{8}{C} \left[ \cos{(\frac{A-B-2C}{2})} - \cos{(\frac{A-B}{2})} \right].
\eea
Thus the terms in $S$ proportional to $\epl{\mu\nu\rh5}$ also cancel.

To summarize thus far, we have shown the cancellation of terms proportional to
$\epl{\mu\nu\rh5}$ and $\epl{\mu\nu\rh6}$ in $S$.  As in the previous
section, this leaves a term proportional to $\epl{\mu\nu\m{12}}$:
\bea
S = \sum_{i=1}^{6} [1.0i] &=& -8 g^5 \int \kt{11}\kt{12} 
\e{ik\times \eta(\frac{1}{2}-\sigma_{12})} \si{\eta}{q} \si{k}{q}
\bisi 
\non \\ & & 
~\propthree \epl{\m{11}\rh1 \rh3} \epl{\rh4 \rh5 \rh6}
\epl{\mu\nu\m{12}} 
\eea
Thus, in analogy with the previous section, we now compute the diagrams in
figure 11, which contain the one-loop vertex correction.  
%%%%% Figure 11 %%%%%
\begin{figure}
\centerline{\epsfxsize=2in\epsfbox{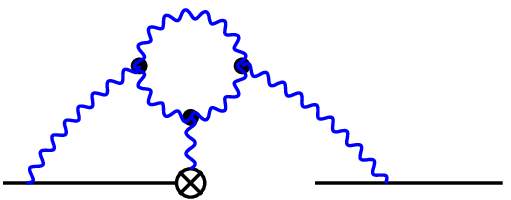} \hspace{1cm}
\epsfxsize=2in\epsfbox{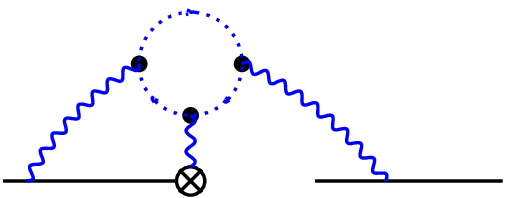} }
\caption{Diagrams [1.07] and [1.08].  Compare with [2.07] and [2.08].}
\end{figure}
Evaluating the diagram with the internal gauge loop we obtain
\bea
[1.07]&=&  (ig)^2 (-2ig)^3 \int \kt{11}\kpt{12} \de{p_{12}+p_{42}-\kp} 
\e{-ip_{12}\cdot\xi_{12}} \e{-\itwo p_{12}\times p_{42}} 
\si{q_1}{q_2} \si{r_1}{r_2} \si{s_1}{s_2} 
\non \\ & &
\epu{\al1 \al2 \al3}
\epu{\bl1 \bl2 \bl3} \epu{\gl1 \gl2 \gl3} {\sst \frac{q_3^{\rh4}
r_3^{\rh5} s_3^{\rh6} p_{11}^{\rh1} p_{12}^{\rh3} p_{42}^{\rh2}}{q_3^2
r_3^2 s_3^2 p_{11}^2 p_{12}^2 p_{42}^2} } (-i p_{42})_{\mu}
\epl{\al3 \rh4 \bl1} \epl{\bl3 \rh5 \gl1} \epl{\gl3 \rh6 \al1}
\epl{\m{11}\rh1 \al2} \epl{\m{12}\rh3 \bl2}
\non \\ & & \epl{\nu\rh2 \gl2}
\de{q_3+r_1} \de{r_3+s_1} \de{s_3+q_1} \de{p_{11}+q_2} \de{p_{12}+r_2}
\de{p_{42}+s_2} - {\sst (\mu\leftrightarrow\nu)}.
\eea
Set $p_{12}=-\eta$, $q_3=-q$, and contract on $\al2, \bl2$ to get:
\bea
[1.07]&=& -8 g^5 \int \kt{11}\kt{12} \e{ik\times \eta
(\frac{1}{2}-\sigma_{12})} \propfour \si{\eta}{q} \si{k}{q} \bisi
\non \\ & &
\epu{\gl1 \gl2 \gl3}
\left[ \epl{\m{11}\rh4 \rh3} \epl{\m{12}\rh5 \gl1} \epl{\gl3 \rh6
\rh1} - \epl{\rh1 \rh4 \rh3} \epl{\m{12}\rh5 \gl1} \epl{\gl3 \rh6
\m{11}} + \epl{\rh1 \rh4 \m{12}} \epl{\rh3 \rh5 \gl1} \epl{\gl3 \rh6
\m{11}} \right]
\non \\ & & ~\times (k-\eta)_{{\sst [}\mu} \epl{\nu {\sst ]} \rh2 \gl2}
\non \\ &=&
-8 g^5 \int \kt{11}\kt{12} \e{ik\times \eta
(\frac{1}{2}-\sigma_{12})} \propfour \si{\eta}{q} \si{k}{q} \bisi
\non \\ & &
\left[ \epl{\m{11}\rh4 \rh3} \epl{\gl3 \rh6 \rh1} \left( \delta^{\gl2
\gl3}_{\m{12}\rh5} - \delta^{\gl2 \gl3}_{\rh5 \m{12}} \right) - 
\epl{\rh1 \rh4 \rh3} \left( \delta^{\gl2 \gl3}_{\m{12}\rh5} -
\delta^{\gl2 \gl3}_{\rh5 \m{12}} \right) \epl{\gl3 \rh6 \m{11}}
\right.
\non \\ & &
~+ \left. \epl{\m{12}\rh1 \rh4} \left( \delta^{\gl1 \gl2}_{\rh6
\m{11}} - \delta^{\gl1 \gl2}_{\m{11}\rh6} \right) \epl{\rh3 \rh5 \gl1}
\right] (k-\eta)_{ {\sst [}\mu} \epl{\nu {\sst ]} \rh2 \gl2}  
\non \\ &=&
-8 g^5 \int \kt{11}\kt{12} \e{ik\times \eta
(\frac{1}{2}-\sigma_{12})} \propfour \si{\eta}{q} \si{k}{q} \bisi
\non \\ & &
\left\{ \left[ \epl{\m{11}\rh4 \rh3}\epl{\rh5 \rh6 \rh1} + \epl{\rh1
\rh3 \rh4} \epl{\rh5 \rh6 \m{11}} + \epl{\m{11}\rh1 \rh4} \epl{\rh5
\rh6 \rh3} \right] (k-\eta)_{ {\sst[}\mu} \epl{ \nu {\sst ]} \rh2
\m{12}} \right.
\non \\ & & ~
\left. - \left[ \epl{\m{11}\rh3 \rh4} \epl{\m{12}\rh1 \rh6} (k-\eta)_{
{\sst [} \mu} \epl{ \nu {\sst ]} \rh2 \rh5} + \epl{\m{12}\rh1 \rh4}
\epl{\m{11}\rh3 \rh5} (k-\eta)_{{\sst [}\mu} \epl{\nu {\sst ]} \rh2
\rh6} \right] \right\}
\non \\ &=&
-8 g^5 \int \kt{11}\kt{12} \e{ik\times \eta
(\frac{1}{2}-\sigma_{12})} \propfour \si{\eta}{q} \si{k}{q} \bisi
\non \\ & &
\bigg\{ \epl{\m{11}\rh1 \rh3} \epl{\rh4 \rh5 \rh6} (k-\eta)_{{\sst
[}\mu} \epl{\nu {\sst ]} \rh2 \m{12}} - \left[ \epl{\m{11}\rh3 \rh4} 
\epl{\m{12}\rh1 \rh6} (k-\eta)_{{\sst [} \mu} \epl{ \nu {\sst ]} \rh2
\rh5} \right.
\non \\ & & ~ 
+ \left. \epl{\m{12}\rh1 \rh4}
\epl{\m{11}\rh3 \rh5} (k-\eta)_{{\sst [}\mu} \epl{\nu {\sst ]} \rh2
\rh6} \right] \bigg\}
\non \\ &=& 
8 g^5 \int \kt{11} \e{ik\times \eta
(\frac{1}{2}-\sigma_{12})} \propthree \si{\eta}{q} \si{k}{q} \bisi
\non \\ & &
\bigg\{ \epl{\m{11}\rh1 \rh3} \epl{\rh4 \rh5 \rh6} \left[ (k\times
\eta) \epl{\mu\nu\rh2} {\sst \frac{(k-\eta)^{\rh2}}{(k-\eta)^2}} 
+ \epl{\mu\nu\m{12}} \kt{12} \right] +  \left[ \epl{\m{11}\rh3 \rh4} 
\epl{\m{12}\rh1 \rh6} (k-\eta)_{{\sst [} \mu} \epl{ \nu {\sst ]} \rh2
\rh5} \right.
\non \\ & & ~ 
~ \left. + \epl{\m{12}\rh1 \rh4}
\epl{\m{11}\rh3 \rh5} (k-\eta)_{{\sst [}\mu} \epl{\nu {\sst ]} \rh2
\rh6} \right] {\sst \frac{(k-\eta)^{\rh2}}{(k-\eta)^2}} \kt{12} \bigg\},
\eea
where we have used (\ref{master identity}) to arrive at the last
line.  The second term cancels the residual piece in $S$.  
Notice that unlike the case of $[2.07]$, the right hand side of
(\ref{master identity}) again survives to create new surface terms
with respect to the $\sigma_{12}$ integration, which we will deal with
after we compute $[1.08]$.

The calculation of the ghost loop graph $[1.08]$ is very similar to
$[2.08]$.  Remembering the minus sign for the closed internal fermion
loop, and the two sets of contractions for the two ghost number flow
directions we have
\bea
[1.08]&=& -(ig)^2 (-2ig)^3 \int \kt{11}\kpt{12} \de{p_{12}+p_{42}-\kp} 
\e{-ip_{12}\cdot\xi_{12}} \e{-\itwo p_{12}\times p_{42}} 
\si{q_1}{q_2} \si{r_1}{r_2} \si{s_1}{s_2} 
\non \\ & &
{\sst (-i)^3 q_1^{\alpha}
r_1^{\beta} s_1^{\gamma} \frac{i^3}{q_3^2 r_3^2 s_3^2}
 \frac{ p_{11}^{\rh1} p_{12}^{\rh3} p_{42}^{\rh2}}
{p_{11}^2 p_{12}^2 p_{42}^2} } (-i p_{42})_{\mu}
\epl{\m{11}\rh1 \alpha} \epl{\m{12}\rh3 \beta} \epl{\nu\rh2 \gamma} 
\de{p_{11}+q_2} \de{p_{12}+r_2} \de{p_{42}+s_2}
\non \\ & & 
\left\{  \de{q_3+r_1}\de{r_3+s_1} \de{s_3+q_1} + \de{q_3+s_1}
\de{s_3+r_1} \de{r_3+q_1} \right\} - {\sst (\mu\leftrightarrow\nu) }.
\eea
Set $p_{12}= -\eta$, and take $r_1=q$ in the first contraction, and
$r_3=q$ in the second to obtain
\bea
[1.08]&=& - 8 g^5 \int \kt{11}\kt{12} \e{ik\times\eta (\frac{1}{2} -
\sigma_{12})} \si{\eta}{q} \si{k}{q} \bisi \propfour
\non \\ & &
\left[ \epl{\m{11}\rh1 \rh6} \epl{\m{12}\rh3 \rh4} (k-\eta)_{ {\sst [}
\mu} \epl{\nu {\sst ]} \rh2 \rh5} + \epl{\m{11}\rh1 \rh4}
\epl{\m{12}\rh3 \rh5} (k-\eta)_{ {\sst [} \mu} \epl{\nu {\sst ]} \rh2
\rh6}
\right].
\eea
Thus the ghost graph $[1.08]$ cancels the last two terms in $[1.07]$.  Adding 
these last two contributions to S, we can summarize our results
thus far as
\bea
\sum_{i=1}^{8} [1.i] &=& 8 g^5 \int \kt{11} 
\e{ik\times \eta(\frac{1}{2}-\sigma_{12})} (k\times \eta) 
\propfour
\non \\ & &
~\times \si{\eta}{q} \si{k}{q} \bisi \epl{\m{11}\rh1 \rh3} \epl{\rh4 \rh5
\rh6} \epl{\mu\nu\rh2}.
\label{one to eight}
\eea
Finally, we must add the contributions shown in figure 12.
%%%%% Figure 12 %%%%%
\begin{figure}
\centerline{\epsfxsize=2in\epsfbox{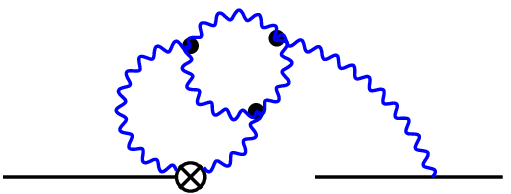} \hspace{1cm}
\epsfxsize=2in\epsfbox{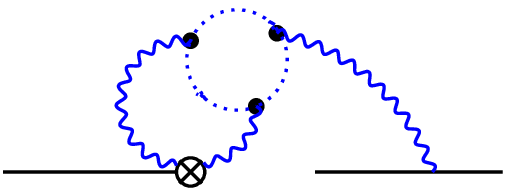} }
\caption{Diagrams [1.09] and [1.10], which cancel the surface terms
generated by [1.07] and [1.08].}
\end{figure}
%%%%%%%%%%%%%%%%%%%%%%
In anticipation of the result we will write $[1.09]$ as follows:
\bea
[1.09]&=& -(ig)^2 (-2ig)^3 \int \kt{11} \de{p_{32}+p_{42}-\kp} 
\e{-\itwo p_{32}\times p_{42}} \si{q_1}{q_2} \si{r_1}{r_2} \si{s_1}{s_2}
\epu{\al1 \al2 \al3}  
\non \\ & &
\epu{\bl1 \bl2 \bl3} \epu{\gl1 \gl2 \gl3}
\epl{\al3 \rh4 \bl1} \epl{\bl3 \rh5 \gl1} \epl{\gl3 \rh6 \al1}
\de{q_3+r_1} \de{r_3+s_1} \de{s_3+q_1} {\sst \frac{ p_{11}^{\rh1}
p_{32}^{\rh3} p_{42}^{\rh2} q_3^{\rh4}
r_3^{\rh5} s_3^{\rh6} }{p_{11}^2 p_{32}^2 p_{42}^2 q_3^2 r_3^2 s_3^2}
}
\non \\ & &
\epl{\m{11}\rh1 \al2}
\de{p_{11}+q_2} \epl{\mu\rh3 \bl2} \epl{\nu\rh2 \gl2} \de{p_{32}+r_2} 
\de{p_{42}+s_2} - {\sst (\mu\leftrightarrow\nu)}
\non \\ &=&
8 ig^5 \int \kt{11} \de{p_{32}+p_{42}-\kp} 
\e{-\itwo p_{32}\times p_{42}} \si{q_1}{q_2} \si{r_1}{r_2} \si{s_1}{s_2}
\epu{\al1 \al2 \al3} \epu{\bl1 \bl2 \bl3} \epu{\gl1 \gl2 \gl3}
\non \\ & &
\epl{\al3 \rh4 \bl1} \epl{\bl3 \rh5 \gl1} \epl{\gl3 \rh6 \al1}
\de{q_3+r_1} \de{r_3+s_1} \de{s_3+q_1} {\sst \frac{ p_{11}^{\rh1}
q_3^{\rh4}
r_3^{\rh5} s_3^{\rh6} }{p_{11}^2 p_{32}^2 p_{42}^2 q_3^2 r_3^2 s_3^2}
} \epl{\m{11} \rh1 \al2} \de{p_{11}+q_2}
\non \\ & &
\frac{1}{2} \left[ {\sst p_{32}^{\rh3} p_{42}^{\rh2} } 
\epl{\mu\rh3 \bl2} \epl{\nu\rh2 \gl2}
\de{p_{32}+r_2} \de{p_{42}+s_2} +
{\sst p_{32}^{\rh2} p_{42}^{\rh3} }
\epl{\mu\rh2 \gl2} \epl{\nu\rh3 \bl2}
\de{p_{32}+s_2} \de{p_{42}+r_2}
\right] 
\non \\ & &
- {\sst (\mu\leftrightarrow\nu) }, 
\eea
where we have written the contractions in two equivalent ways in the
second line.  They are equivalent because the
contractions of the two commutator gauge field sources into the two
remaining (i.e. after the contraction from the source on $W(k)$ into
an internal vertex is fixed) internal vertices are arbitrary:  we are 
averaging again by
pre-emptively performing the variable change $\eta\rightarrow k-\eta$
in the second term. 

Then choosing $q_3=-q$, and $p_{32}=-\eta$ in the first term, 
$p_{42}=-\eta$ in the second, and contracting on $\al3, \bl3, \gl3$ we obtain
\bea
[1.09]&=& 4i g^5 \int \kt{11} \propfour \si{\eta}{q} \si{k}{q} \bisi
\epu{\al1 \al2 \al3} \epu{\bl1 \bl2 \bl3}
\non \\ & &
\epu{\gl1 \gl2 \gl3}
\epl{\al3 \rh4 \bl1} \epl{\bl3 \rh5 \gl1} \epl{\gl3 \rh6 \al1} 
\epl{\m{11}\rh1 \al2}
\left[
\e{\itwo k\times \eta} \epl{\mu \rh3 \bl2} \epl{\nu \rh2 \gl2} +
\e{-\itwo k\times \eta} \epl{\mu\rh2 \gl2} \epl{\nu\rh3 \bl2} - {\sst
(\mu\leftrightarrow\nu)} \right]
\non \\&=&
-8 g^5 \int \kt{11} \propfour \si{\eta}{q} \si{k}{q} \bisi \si{k}{\eta}
\epu{\al1 \al2 \al3}
\non \\ & &
\epu{\bl1 \bl2 \bl3} \epu{\gl1 \gl2 \gl3} 
\epl{\al3 \rh4 \bl1} \epl{\bl3 \rh5 \gl1} \epl{\gl3 \rh6 \al1}
\epl{\m{11}\rh1 \al2}
\left[ \epl{\mu\nu\rh2} \epl{\rh3 \bl2 \gl2} - \epl{\rh3 \bl2 \rh2}
\epl{\mu\nu\gl2} \right]
\non \\ &=&
-8 g^5 \int \kt{11} \propfour \si{\eta}{q} \si{k}{q} \bisi \si{k}{\eta}
\non \\ & &
{\sst \left[ \delta^{\al2 \bl2 \gl2}_{\rh5 \rh6 \rh4} - \delta^{\al2 \bl2
\gl2}_{\rh5 \rh4 \rh6} - \delta^{\al2 \bl2 \gl2}_{\rh6 \rh5 \rh4} -
\delta^{\al2 \bl2 \gl2}_{\rh4 \rh6 \rh5} \right] }
\epl{\m{11}\rh1 \al2}
\left[ \epl{\mu\nu\rh2} \epl{\rh3 \bl2 \gl2} - \epl{\rh3 \bl2 \rh2}
\epl{\mu\nu\gl2} \right]
\non \\ &=&
-8 g^5 \int \kt{11} \propfour \si{\eta}{q} \si{k}{q} \bisi \si{k}{\eta}
\non \\ & &
\left\{ \epl{\mu\nu\rh2} \left[ \epl{\m{11}\rh1 \rh5} \epl{\rh3 \rh6
\rh4} - \epl{\m{11}\rh1 \rh5} \epl{\rh3 \rh4 \rh6} - \epl{\m{11}\rh1
\rh6} \epl{\rh3 \rh5 \rh4} - \epl{\m{11}\rh1 \rh4} \epl{\rh3 \rh6
\rh5} \right] \right.
\non \\ & &
~- \left[ \epl{\m{11}\rh1 \rh5} \epl{\rh3 \rh6 \rh2} \epl{\mu\nu\rh4} -
\epl{\m{11}\rh1 \rh5} \epl{\rh3 \rh4 \rh2} \epl{\mu\nu\rh6} -
\epl{\m{11}\rh1 \rh6} \epl{\rh3 \rh5 \rh2} \epl{\mu\nu\rh4} -
\right.
\non \\ & &
\left. \left. -
\epl{\m{11}\rh1 \rh4} \epl{\rh3 \rh6 \rh2} \epl{\mu\nu\rh5} 
\right] \right\}
\non \\ &=&
-8 g^5 \int \kt{11} \propfour \si{\eta}{q} \si{k}{q} \bisi \si{k}{\eta}
\non \\ & &
\left\{ \epl{\mu\nu\rh2} \epl{\m{11}\rh1 \rh5} \epl{\rh3 \rh6 \rh4} +
\epl{\mu\nu\rh2} \epl{\m{11}\rh1 \rh3} \epl{\rh4 \rh5 \rh6} +
\epl{\mu\nu\rh4} \epl{\m{11}\rh1 \rh2} \epl{\rh3 \rh5 \rh6} \right.
\non \\ & &
\left.
~ + \epl{\mu\nu\rh6} \epl{\m{11}\rh1 \rh5} \epl{\rh2 \rh3 \rh4} +
\epl{\mu\nu\rh5} \epl{\m{11}\rh1 \rh4} \epl{\rh2 \rh3 \rh6} \right\}
\non \\ &=&
-8 g^5 \int \kt{11} \propfour \si{\eta}{q} \si{k}{q} \bisi \si{k}{\eta}
\non \\ & &
\left\{ \epl{\mu\nu\rh2} \epl{\m{11}\rh1 \rh3} \epl{\rh4 \rh5 \rh6} +
\epl{\mu\nu\rh3} \epl{\m{11}\rh1 \rh5} \epl{\rh2 \rh6 \rh4} +
\epl{\mu\nu\rh4}  \epl{\m{11}\rh1 \rh5} \epl{\rh2\rh3\rh6} 
\right.
\non \\ & & 
~ \left. + \epl{\mu\nu\rh4} \epl{\m{11}\rh1 \rh2} \epl{\rh3 \rh5 \rh6} +
\epl{\mu\nu\rh5} \epl{\m{11}\rh1 \rh4} \epl{\rh2 \rh3 \rh6} \right\}
\non \\ &=&
-8 g^5 \int \kt{11} \propfour \si{\eta}{q} \si{k}{q} \bisi \si{k}{\eta}
\non \\ & &
\left\{ \epl{\mu\nu\rh2} \epl{\m{11}\rh1 \rh3} \epl{\rh4 \rh5 \rh6} +
\epl{\mu\nu\rh3} \epl{\m{11}\rh1 \rh2} \epl{\rh4 \rh5 \rh6} +
\epl{\mu\nu\rh3} \epl{\m{11}\rh1 \rh6} \epl{\rh2 \rh5 \rh4} \right.
\non \\ & & ~
\left. + \epl{\mu\nu\rh4} \epl{\m{11}\rh1 \rh6} \epl{\rh2 \rh3 \rh5}  
+ \epl{\mu\nu\rh5} \epl{\m{11}\rh1 \rh4} \epl{\rh2 \rh3 \rh6} \right\},
\label{[1.09]}
\eea
where we have repeatedly applied (\ref{epsilon identity 1}) in
conjunction with the following momentum identity:
\be
\epl{\rh2 \rh5 \rh6} (k-\eta)^{\rh2} (q+\eta)^{\rh5} (k+q)^{\rh6} = 0.
\label{momentum identity}
\ee

Finally we compute the ghost graph $[1.10]$ paired with
$[1.09]$, and show that it cancels the last three terms in (\ref{[1.09]}):
\bea
[1.10]&=& + (ig)^2 (-2ig)^3 \int \kt{11} \de{p_{32}+p_{42}-\kp}
\e{-\itwo p_{32}\times p_{42}} \si{q_1}{q_2} \si{r_1}{r_2}
\si{s_1}{s_2} {\sst (-i)^3 q_1^{\alpha}
r_1^{\beta} s_1^{\gamma} }
\non \\ & &
\times 
{\sst \frac{i^3}{q_3^2 r_3^2 s_3^2}
 \frac{ p_{11}^{\rh1} p_{12}^{\rh3} p_{42}^{\rh2} }
{p_{11}^2 p_{12}^2 p_{42}^2} } \epl{\m{11}\rh1 \alpha}
\epl{\mu\rh3 \beta} \epl{\nu\rh2 \gamma} \de{p_{11}+q_2}
\de{p_{32}+r_2} \de{p_{42}+s_2} 
\non \\ & &
\times \left[ \de{q_3+r_1}
\de{r_3+s_1} \de{s_3+q_1} + \de{q_3+s_1} \de{s_3+r_1} \de{r_3+q_1} \right]  
- {\sst (\mu\leftrightarrow\nu)}
\non \\ &=&
-4 i g^5 \int \kt{11} \de{p_{32}+p_{42}-\kp}
\e{-\itwo p_{32}\times p_{42}} \si{q_1}{q_2} \si{r_1}{r_2}
\si{s_1}{s_2} {\sst \frac{ q_1^{\alpha}
r_1^{\beta} s_1^{\gamma} p_{11}^{\rh1} }{q_3^2 r_3^2 s_3^2 p_{11}^2 }
}
\non \\ & &
\epl{\m{11}\rh1 \alpha} \de{p_{11}+q_2} \left\{ {\sst
\frac{p_{32}^{\rh3} p_{42}^{\rh2}}{p_{32}^2 p_{42}^2} } \epl{\mu\rh3
\beta} \epl{\nu\rh2 \gamma} \de{p_{32}+r_2} \de{p_{42}+s_2} 
\bigg[ \de{q_3+r_1} \de{r_3+s_1} \de{s_3+q_1} \right.
\non \\ & &
~~  + \de{q_3+s_1} \de{s_3+r_1} \de{r_3+q_1} \bigg] + {\sst
\frac{p_{32}^{\rh2} p_{42}^{\rh3}}{p_{32}^2 p_{42}^2} } \epl{\mu\rh2
\gamma} \epl{\nu\rh3 \beta} \de{p_{32}+s_2} \de{p_{42}+r_2} \times
\non \\ & &
~\times \left.
 \bigg[ \de{q_3+r_1} \de{r_3+s_1} \de{s_3+q_1} + 
 +  \de{q_3+s_1} \de{s_3+r_1} \de{r_3+q_1} \bigg] 
\right\} - {\sst (\mu\leftrightarrow\nu)},
\eea
where again we have written each of the terms in the first line
in two different ways, and averaged their contributions.  Thus for the
first two terms take $p_{32}=-\eta$, and for the second two
$p_{42}=-\eta$.  Furthermore take $r_1=q$ for the first and third
terms, and $r_3=q$ for the second and fourth terms.  Then we have
\bea
[1.10] &=& 4i g^5 \int \kt{11} \si{\eta}{q} \si{k}{q} \bisi \propfour
\non \\ & &
\left\{ \e{\itwo k\times \eta} \left[ \epl{\m{11}\rh1 \rh6} \epl{\mu\rh3
\rh4} \epl{\nu \rh2 \rh5} + \epl{\m{11}\rh1 \rh4} \epl{\mu\rh3 \rh5}
\epl{\nu\rh2 \rh6} \right] \right. 
\non \\ & &
\left.
~~ + \e{-\itwo k\times \eta} \left[ 
\epl{\m{11}\rh1 \rh6} \epl{\mu\rh2
\rh5} \epl{\nu \rh3 \rh4} + \epl{\m{11}\rh1 \rh4} \epl{\mu\rh2 \rh6}
\epl{\nu\rh3 \rh5}  \right] \right\} - {\sst (\mu\leftrightarrow\nu)}
\non \\ &=&
8 g^5 \int \kt{11} \si{\eta}{q} \si{k}{q} \bisi \si{k}{\eta}
\propfour
\non \\ & &
\left\{ \epl{\m{11}\rh1 \rh6} \left[ \epl{\rh5 \rh2 \nu} \epl{\mu \rh3
\rh4} - \epl{\rh5 \rh2 \mu} \epl{\nu \rh3 \rh4} \right] + 
\epl{\m{11}\rh1 \rh4} \left[ \epl{\rh6 \rh2 \nu} \epl{\mu \rh3 \rh5} 
- \epl{\rh6 \rh2 \mu} \epl{\nu \rh3 \rh5} \right] 
\right\}
\non \\ &=&
8 g^5 \int \kt{11} \si{\eta}{q} \si{k}{q} \bisi \si{k}{\eta}
\propfour
\non \\ & &
\left\{ \epl{\m{11}\rh1 \rh6} \left[ \epl{\rh2 \rh3 \rh5}
\epl{\mu\nu\rh4} + \epl{\rh2 \rh5 \rh4} \epl{\mu\nu\rh3} \right] +
\epl{\m{11}\rh1 \rh4} \epl{\rh2 \rh3 \rh6} \epl{\mu\nu\rh5} \right\},
\eea
where we have used (\ref{epsilon identity 2}) and the momentum
identity (\ref{momentum identity}).  Thus the sum of the diagrams in
figure 12 is given by
\bea
\sum_{i=9}^{10} [1.i] &=& 
-8 g^5 \int \kt{11} \propfour \si{\eta}{q} \si{k}{q} \bisi \si{k}{\eta}
\non \\ & &
\left\{ \epl{\mu\nu\rh2} \epl{\m{11}\rh1 \rh3} \epl{\rh4 \rh5 \rh6} +
\epl{\mu\nu\rh3} \epl{\m{11}\rh1 \rh2} \epl{\rh4 \rh5 \rh6} \right\}.
\eea
It turns out that the two terms in brackets are the same.  To see
this make the following changes of variables in the second term:  
$\eta \rightarrow k-\eta$, and $q \rightarrow -(k+q)$.  We
obtain the identity
\bea
& & \int \si{k}{\eta} \si{\eta}{q} \bisi \si{k}{q}
\propfour
\epl{\mu\nu\rh3} \epl{\m{11}\rh1 \rh2} \epl{\rh4 \rh5 \rh6}
\non \\ &=&
\int \si{k}{\eta} \bisi \si{\eta}{q} \si{k}{q}
\propfour
\epl{\mu\nu\rh2} \epl{\m{11}\rh1 \rh3} \epl{\rh4 \rh5 \rh6}.
\non \\
\eea

Thus comparing with (\ref{one to eight}), we obtain finally
\bea
\sum_{i=9}^{10} [1.i] &=& -16 g^5 \int \propfour \si{\eta}{q} \si{k}{q} 
\bisi \si{k}{\eta} 
\non \\ & & ~\times 
\epl{\mu\nu\rh2} \epl{\m{11}\rh1 \rh3} \epl{\rh4 \rh5 \rh6}
\non \\ &=& - \sum_{i=1}^{8} [1.i],
\eea
after performing the trivial $\sigma_{12}$ integration in (\ref{one to eight}).
That is
\be
\sum_{i=1}^{10} [1.i] = 0.
\ee

Except for the two-loop propagator corrections, this completes the
contributions from $O(g)$ terms in $W(k)$ to the correlator $\vev{W(k)
O_{\mu\nu}(\kp)}$.  The set of one-particle irreducible diagrams
contributing to the two-loop propagator corrections which do
not themselves contain one-loop propagator corrections within them are
shown in figure 13.  Each of
these graphs will have the same noncommutative
phases and propagators, and we will demonstrate that the tensor
algebra arranges a cancellation amongst them.  This does not depend on
the Wilson line structure, and in particular on the identity
(\ref{master identity}) that has been involved throughout these
computations thus far.

%%%%% Figure 13 %%%%%
\begin{figure}
\centerline{\epsfxsize=2in\epsfbox{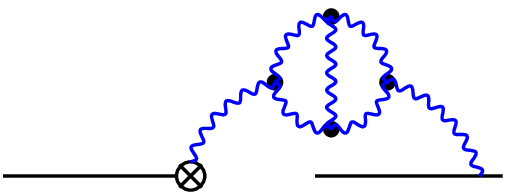} \hspace{1cm}
\epsfxsize=2in\epsfbox{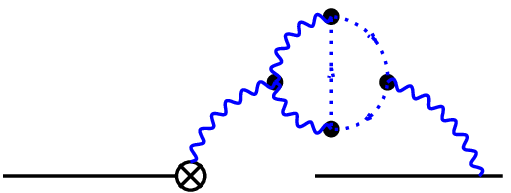} }
\centerline{\epsfxsize=2in\epsfbox{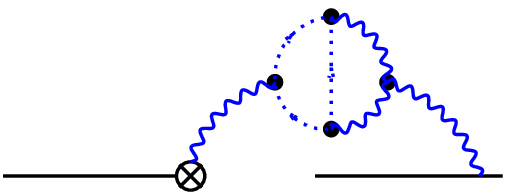} \hspace{1cm}
\epsfxsize=2in\epsfbox{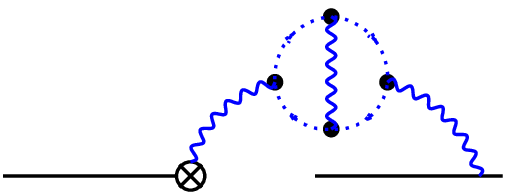} }
\caption{Intrinsically two-loop, one-particle irreducible propagator correction
contributions to $\vev{W(k) O_{\mu\nu}(\kp)}$.  Diagrams [1.11]-[1.14]
respectively.} 
\end{figure}

The most complicated is the pure gauge loop, diagram $[1.11]$:
\bea
[1.11]&=& (ig)(-2ig)^4 \int \kt{11} \de{p_{42}-\kp}
\si{q_1}{q_2} \si{r_1}{r_2} \si{s_1}{s_2} \si{t_1}{t_2}
\epu{\al1 \al2 \al3} \epu{\bl1 \bl2 \bl3}
\non \\ & &
\epu{\gl1 \gl2 \gl3}
\epu{\delta_1 \delta_2 \delta_3} {\sst \frac{ p_{11}^{\rh1}
p_{42}^{\rh7} q_1^{\rh4} q_3^{\rh6} r_2^{\rh5} r_3^{\rh2} s_1^{\rh3}}{
p_{11}^2 p_{42}^2 q_1^2 q_3^2 r_2^2 r_3^2 s_1^2} } 
(-ip_{42})_{{\sst [} \mu}
\epl{\nu {\sst ]} \rh7 \delta_2} \epl{\m{11} \rh1 \al2}
\epl{\al1 \rh4 \gl3}
\epl{\al3 \rh6 \bl1} \epl{\bl2 \rh5 \gl2} \epl{\bl3 \rh2 \delta_1}
\non \\ & & 
\epl{\gl1 \rh3 \delta_3}~
\de{p_{11}+q_2} \de{p_{42}+ t_2} \de{q_1+s_3} \de{q_3+r_1}
\de{r_2+s_2} \de{r_3+t_1} \de{s_1+t_3}
\eea 
Setting $s_3=q$, and $s_1=\eta$, and contracting on $\al1, \bl1,
\gl1$, and $\delta_1$ we obtain
\bea
[1.11]&=& 16 g^5 \int \kt{11} \si{k}{q} \bisi \si{\eta}{q} \si{k}{\eta} {\sst
\frac{k^{\rh1} (k-\eta)^{\rh2} \eta^{\rh3} q^{\rh4} (q+\eta)^{\rh5}
(k+q)^{\rh6} k^{\rh7} }{k^2 (k-\eta)^2 \eta^2 q^2 (q+\eta)^2 (k+q)^2
k^2}}
\non \\ & &
\times k_{\mu} \left[ 2 \epl{\m{11}\rh1 \rh4} \epl{\nu \rh7 \rh6} \epl{\rh2
\rh5 \rh3} + 2 \epl{\m{11}\rh1 \rh4} \epl{\nu \rh7 \rh2} \epl{\rh3
\rh5 \rh6} + 2 \epl{\m{11}\rh1 \rh3} \epl{\nu \rh7 \rh2} \epl{\rh6
\rh5 \rh4} \right.
\non \\ & &
~~ \left. + \left( \epl{\m{11}\rh1 \rh3} \epl{\nu \rh7 \rh6} \epl{\rh4 \rh5
\rh2} + \epl{\m{11} \rh1 \rh2} \epl{\nu \rh7 \rh4} \epl{\rh6 \rh5
\rh3} \right)\right] - {\sst (\mu\leftrightarrow\nu)}
\non \\ &=&
32 g^5 \int \kt{11} \si{k}{q} \bisi \si{\eta}{q} \si{k}{\eta} {\sst
\frac{k^{\rh1} (k-\eta)^{\rh2} \eta^{\rh3} q^{\rh4} (q+\eta)^{\rh5}
(k+q)^{\rh6} k^{\rh7} }{k^2 (k-\eta)^2 \eta^2 q^2 (q+\eta)^2 (k+q)^2
k^2}}
\non \\ & &
\times k_{\mu} \left[ \epl{\m{11}\rh1 \rh4} \epl{\nu \rh7 \rh6} \epl{\rh2
\rh5 \rh3} + \epl{\m{11}\rh1 \rh5} \epl{\nu \rh7 \rh2} \epl{\rh4
\rh6 \rh3} \right] - {\sst (\mu\leftrightarrow\nu)},
\eea
where in addition to (\ref{epsilon identity 2}) and 
(\ref{momentum identity}), we have used
\be
\epl{\rh3 \rh4 \rh5} \eta^{\rh3} q^{\rh4} (q+\eta)^{\rh5} = 0,
\label{momentum identity 2}
\ee
and changed variables $\eta \rightarrow k-\eta, q \rightarrow -(k+q)$
in the last term on the first line to cancel the penultimate term
there.  Diagram $[1.12]$ is given by
\bea
[1.12]&=& -ig (-2ig)^4 \int \kt{11} \de{p_{42}-\kp} \si{q_1}{q_2} 
\si{r_1}{r_2} \si{s_1}{s_2} \si{t_1}{t_2} \epu{\delta_1 \delta_2 \delta_3}
{\sst \frac{ p_{11}^{\rh1} p_{42}^{\rh7}
r_2^{\rh2} s_2^{\rh3} i^3}{p_{11}^2 p_{42}^2 r_2^2 s_2^2 q_3^2 r_3^2 s_3^2} }
\non \\ & &
(-i)^3 q_1^{\al2} r_1^{\bl2} s_1^{\gl2} (-ip_{42})_{[ \mu} \epl{\nu ]
\rh7 \delta_2} \epl{\m{11}\rh1 \al2} \epl{\bl2 \rh2 \delta_1}
\epl{\gl2 \rh3 \delta_3} \de{p_{42}+t_2} \de{p_{11}+q_2} \de{r_2+t_1}
\non \\ & &
\de{s_2+t_3} \left[ \de{q_3+r_1} \de{r_3+s_1} \de{s_3+q_1} +
\de{q_3+s_1} \de{s_3+r_1} \de{r_3+q_1} \right]. 
\eea
In the first term set $s_2=\eta, s_3=q$, and in the second set
$s_2=\eta, s_1=q$.  Then contract on $\delta_1$ to obtain:
\bea
[1.12] &=& -16 g^5 \int \kt{11} \si{k}{q} \bisi \si{\eta}{q} \si{k}{q} 
{\sst
\frac{k^{\rh1} (k-\eta)^{\rh2} \eta^{\rh3} q^{\rh4} (q+\eta)^{\rh5}
(k+q)^{\rh6} k^{\rh7} }{k^2 (k-\eta)^2 \eta^2 q^2 (q+\eta)^2 (k+q)^2
k^2}}
\non \\ & &
\times k_{\mu} \left[ \epl{\m{11} \rh1 \rh4} \left( \epl{\nu\rh7 \rh6}
\epl{\rh5 \rh3 \rh2} - \epl{\nu \rh7 \rh2} \epl{\rh5 \rh3 \rh6}
\right) + \epl{\m{11}\rh1 \rh6} \epl{\rh4 \rh3 \rh2} \epl{\nu \rh7
\rh5}
\right] - {\sst (\mu\leftrightarrow\nu)}
\non \\ &=& 
-32 g^5 \int \kt{11} \si{k}{q} \bisi \si{\eta}{q} \si{k}{q} 
{\sst
\frac{k^{\rh1} (k-\eta)^{\rh2} \eta^{\rh3} q^{\rh4} (q+\eta)^{\rh5}
(k+q)^{\rh6} k^{\rh7} }{k^2 (k-\eta)^2 \eta^2 q^2 (q+\eta)^2 (k+q)^2
k^2}}
\non \\ & &
\times k_{\mu} \epl{\m{11} \rh1 \rh4} \left[  \epl{\nu\rh7 \rh6}
\epl{\rh5 \rh3 \rh2} - \epl{\nu \rh7 \rh2} \epl{\rh5 \rh3 \rh6}
\right] - {\sst (\mu\leftrightarrow\nu) },
\eea
where we have used (\ref{epsilon identity 2}) several times in
conjunction with (\ref{momentum identity 2}), and
\be
\epl{\rh1 \rh2 \rh3} k^{\rh1} (k-\eta)^{\rh2} \eta^{\rh3} = \epl{\rh1
\rh4 \rh6} k^{\rh1} q^{\rh4} (k+q)^{\rh6} = 0.
\ee
Diagram $[1.13]$ evaluates similarly:
\bea
[1.13] &=& -ig (-2ig)^4 \int \kt{11} \de{p_{42}-\kp} \si{q_1}{q_2} 
\si{r_1}{r_2} \si{s_1}{s_2} \si{t_1}{t_2} \epu{\al1 \al2 \al3}
{\sst \frac{ p_{11}^{\rh1} p_{42}^{\rh7}
r_2^{\rh2} s_2^{\rh3} i^3}{p_{11}^2 p_{42}^2 r_2^2 s_2^2 q_3^2 r_3^2
s_3^2} }
\non \\ & &
(-i)^3 t_1^{\delta_2} r_1^{\bl2} s_1^{\gl2} (-ip_{42})_{[ \mu}
\epl{\nu ] \rh7 \delta_2} \epl{\m{11}\rh1 \al2} \epl{\bl2 \rh4 \al1}
\epl{\gl2 \rh6 \al3} \de{p_{11}+q_2} \de{p_{42}+t_2} \de{r_2+q_1}
\non \\ & &
\de{s_2+q_3} \left[ \de{r_3+s_1} \de{s_3+t_1} \de{t_3+r_1} +
\de{r_3+t_1} \de{t_3+s_1} \de{s_3+r_1}  \right].
\eea
Take $r_2=q, r_1=\eta$ in the first term, and $r_2=q, r_3=\eta$ in the
second, and contract on $\al1$ to get
\bea
[1.13]&=& 16 g^5 \int \kt{11} \si{k}{q} \si{\eta}{q} \bisi
\si{k}{\eta} {\sst
\frac{k^{\rh1} (k-\eta)^{\rh2} \eta^{\rh3} q^{\rh4} (q+\eta)^{\rh5}
(k+q)^{\rh6} k^{\rh7} }{k^2 (k-\eta)^2 \eta^2 q^2 (q+\eta)^2 (k+q)^2
k^2}}
\non \\ & &
k_{\mu} \left[ \epl{\m{11}\rh1 \rh3} \epl{\nu \rh7 \rh2} \epl{\rh5
\rh6 \rh4} - \epl{\m{11}\rh1 \rh4} \epl{\nu \rh7 \rh2} \epl{\rh5 \rh6
\rh3} + \epl{\m{11} \rh1 \rh5} \epl{\nu \rh7 \rh3} \epl{\rh2 \rh6
\rh4} \right] - {\sst (\mu\leftrightarrow\nu)}
\non \\ &=& 
16 g^5 \int \kt{11} \si{k}{q} \si{\eta}{q} \bisi
\si{k}{\eta} {\sst
\frac{k^{\rh1} (k-\eta)^{\rh2} \eta^{\rh3} q^{\rh4} (q+\eta)^{\rh5}
(k+q)^{\rh6} k^{\rh7} }{k^2 (k-\eta)^2 \eta^2 q^2 (q+\eta)^2 (k+q)^2
k^2}}
\non \\ & &
\times
k_{\mu} \left[ \epl{\nu\rh7 \rh2} \epl{\m{11}\rh1 \rh5} \epl{\rh3 \rh6
\rh4}
+ \epl{\nu\rh7 \rh3} \epl{\m{11}\rh1 \rh5} \epl{\rh2 \rh6 \rh4} \right]
- {\sst (\mu\leftrightarrow\nu)}
\non \\ &=&
-32 g^5 \int \kt{11} \si{k}{q} \si{\eta}{q} \bisi
\si{k}{\eta} {\sst
\frac{k^{\rh1} (k-\eta)^{\rh2} \eta^{\rh3} q^{\rh4} (q+\eta)^{\rh5}
(k+q)^{\rh6} k^{\rh7} }{k^2 (k-\eta)^2 \eta^2 q^2 (q+\eta)^2 (k+q)^2
k^2}}
\non \\ & &
\times k_{ {\sst [}\mu } \epl{\nu {\sst ]} \rh7 \rh2} \epl{\m{11}\rh1 \rh5} 
\epl{\rh3 \rh4 \rh6}.
\eea
This leaves finally the ghost loop $[1.14]$:
\bea
[1.14] &=& -ig (-2ig)^4 \int \kt{11} \de{p_{42}-\kp} \si{q_1}{q_2} 
\si{r_1}{r_2} \si{s_1}{s_2} \si{t_1}{t_2} 
{\sst \frac{ p_{11}^{\rh1} p_{42}^{\rh7}
r_2^{\rh5} i^4}{p_{11}^2 p_{42}^2 r_2^2  q_3^2 r_3^2
s_3^2 t_3^2} } 
\non \\ & &
(-ip_{42})_{{\sst[}\mu} \epl{\nu {\sst]} \rh7 \delta_2}
\epl{\m{11}\rh1 \al2} \epl{\bl2 \rh5 \gl2} (-i)^4 q_1^{\al2}
r_1^{\bl2} s_1^{\gl2} t_1^{\delta_2} \de{p_{42}+t_2} \de{p_{11}+q_2}
\de{r_2+s_2} 
\non \\ & &
\times \left[ \de{q_3+r_1} \de{r_3+t_1} \de{t_3+s_1} \de{s_3+q_1} +
\de{q_3+s_1} \de{s_3+t_1} \de{t_3+r_1} \de{r_3+q_1}
\right].
\non \\
\eea
Taking $s_1= \eta, s_3=q$ in the first term, and $s_1= q, s_3=\eta$ in
the second term, we obtain
\bea
[1.14]&=& -16 g^5 \int \kt{11} \si{k}{q} \bisi \si{\eta}{q}
\si{k}{\eta} {\sst \frac{k^{\rh1} (k-\eta)^{\rh2} \eta^{\rh3} q^{\rh4} 
(q+\eta)^{\rh5} (k+q)^{\rh6} k^{\rh7} }
{k^2 (k-\eta)^2 \eta^2 q^2 (q+\eta)^2 (k+q)^2 k^2} }
\non \\ & &
\times k_{\mu} \left[ \epl{\m{11}\rh1 \rh4} \epl{\rh6 \rh5 \rh3}
\epl{\nu \rh7 \rh2} + \epl{\m{11}\rh1 \rh6} \epl{\rh2 \rh5 \rh4}
\epl{\nu \rh7 \rh3} \right] - {\sst (\mu\leftrightarrow\nu)}
\non \\ &=&
-32 g^5 \int \kt{11} \si{k}{q} \bisi \si{\eta}{q}
\si{k}{\eta} {\sst \frac{k^{\rh1} (k-\eta)^{\rh2} \eta^{\rh3} q^{\rh4} 
(q+\eta)^{\rh5} (k+q)^{\rh6} k^{\rh7} }
{k^2 (k-\eta)^2 \eta^2 q^2 (q+\eta)^2 (k+q)^2 k^2} }
\non \\ & &
\times k_{ {\sst [} \mu} \epl{\nu {\sst ]} \rh7 \rh2}
\epl{\m{11}\rh1 \rh4} \epl{\rh6 \rh5 \rh3},
\eea
using (\ref{epsilon identity 2}) repeatedly, or equivalently changing variables
$\eta \rightarrow (k-\eta), q \rightarrow -(k+q)$ in the second term,
which again effectively interchanges the $\rh2 \leftrightarrow \rh3$ and
$\rh4 \leftrightarrow \rh6$ indices.  

Comparing the results from $[1.11]-[1.14]$, we see that
\be
\sum_{i=11}^{14} [1.i] = 0.
\ee
The remaining contributions to the two-loop propagator all involve the
linearly divergent one-loop corrections to the gauge or ghost
propagator as subgraphs, and either formally cancel pairwise or
vanish, or need to be carefully regulated as per our discussion in
section 2, and that in
\cite{Kaminsky:2003}.  As discussed in the latter, the
tensor structure and the momentum dependence of the propagators are not
modified by their one-loop corrections.  Thus the presence of them as 
subgraphs in the remaining two-loop graphs, will essentially
reduce these two-loop graphs to the one-loop propagator correction
case.  Since the one-loop corrections to the propagator occurred in
our calculation as a quantum correction to the single nonvanishing graph at 
tree-level that reflects the equivalence between the commutative and
noncommutative theories, and the net effect of these
corrections to our calculation was to induce a harmless
finite renormalization to the Seiberg-Witten map itself (if the
equivalence is to be maintained), any such effect at two-loops can be 
similarly absorbed into a two-loop renormalization of 
Seiberg-Witten map.  We will not consider them further here.

To summarize this section, we conclude that the sum of the $O(g)$ 
contributions from $W(k)$ to the correlator $\vev{W(k)
O_{\mu\nu}(\kp)}$ at $O(g^5)$ vanishes.

%%%%% Section 6 %%%%%
\section{Discussion and Conclusions}
\setcounter{equation}{0}

To summarize the results from sections 3 to 5, we have demonstrated
the complete cancellation amongst all of the
order $g^5$ contributions to the correlator $\vev{W(k)
O_{\mu\nu}(\kp)}$.  Thus far our discussion has been almost entirely
formal to keep the calculations and the cancellations
transparent, and so let us now discuss the issue of regularization.  In
\cite{Kaminsky:2003} we showed that a point-splitting regulator
separating gauge field sources on the Wilson lines or the gauge
fields of the field strength commutator, and natural \cite{Okawa:2000}
from the point of view of the computation of disk amplitudes in string theory
where noncommutative gauge theory in spacetime arises from a
point-splitting regularization of operators on the worldsheet
boundary \cite{Seiberg:1999}, was sufficient to regularize
the divergences arising from the graphs studied therein.
Specifically, the path parameter integrations along the Wilson lines
had their integration ranges restricted so that operators were never
closer than $\epsilon \tilde{k}$ together.  The operators making up
the commutator
in the field strength were similarly separated from each other, and in
the case of one graph, from an operator along the path-ordered
exponential part of $O_{\mu\nu}(k)$.  In the na\"{i}ve limit $\epsilon
\rightarrow 0$, the unregularized composite operators are recovered.  The
essential effect of this regulator is to ensure that the noncommutative
phases do not vanish in each graph: putative planar components that
are superficially divergent are regulated by noncommutative phases of
the form $\e{i a \epsilon k\times p}$, where $p$ is a loop momentum,
and $a$ is some half-integer.  It has the explicit advantage
of not modifying the propagators or vertices, and so the tensor algebra in
each of the graphs is unaltered.  The expense paid is that the 
noncommutative phases
between graphs become somewhat complicated relative to each other, and
graphs that formally cancel now differ by terms of order $\epsilon$.  
Furthermore, it obviously cannot regulate divergences coming from graphs with
loops that are internal (i.e. do not connect to the Wilson lines themselves).

This regulator can be naturally extended to the case
where there are three or more sources along $W(k)$, as well as the
case where there are two sources on $O_{\mu\nu}(\kp)$ in addition to
those from the commutator.  For example we take
\be
{\sst \int_0^1 d\sigma_1 \int_0^{\sigma_1} d\sigma_2 A(x+\xi(\sigma_1)) \ast
A(x+\xi(\sigma_2)) \ast A(x) \ast \e{i k\cdot x} \rightarrow
\int_{\epsilon}^{1-\epsilon} d\sigma_1
\int_{2\epsilon}^{\sigma_1-\epsilon} d\sigma_2 A(x+\xi(\sigma_1)) \ast
A(x+\xi(\sigma_2)) \ast A(x) \ast \e{i k\cdot x} },
\ee
where both upper and lower limits of integration are modified
because of (\ref{noncommutative identity}), and the cyclicity property
of the star product under the integration over spacetime.  As long as
the operators are separated
along the Wilson lines, a nonvanishing noncommutative phase will 
protect loop momenta integrations associated with gauge sources along $W(k)$ or
$O_{\mu\nu}(\kp)$ carrying those momenta.  

It should be clear
that the computations in $[3.01]-[3.02]$, and $[2.01]-[2.03]$, being 
essentially identical to those in \cite{Kaminsky:2003}, will go through
rigourously with this regulator.  The graphs $[2.04]-[2.06]$,
which are crossings of those in $[2.01]-[2.03]$, 
should also be made rigourously finite, so that we can legitimately sum those
contributions and apply (\ref{master identity}).  The $q$ integration
in that sum is finite by power
counting since $\epl{\rh4 \rh5 \rh6} q^{\rh4} (q+\eta)^{\rh5}
(k+q)^{\rh6}$ is linear in $q$, and the (subsequent) $\eta$
integration is protected by a nonvanishing noncommutative phase with
the regulator applied.  On the other hand,
$[2.07]$ and $[2.08]$ contain, in addition, logarithmically divergent
terms with
respect to the $q$ integration: the integration associated with the
internal loop in those graphs.  As discussed above, these graphs, which
contain the one-loop vertex corrections as subgraphs, need to be
regulated separately.  The similarity between graphs $[2.06]$ and
$[2.07]$ however, suggests the existence of a regulator which would
retain the compatibility
between $[2.04]-[2.06]$ and $[2.07]-[2.08]$, and make the final
cancellation with $\sum_{i=4}^6 [2.i]$ more rigourous.  Specifically
we might regulate $[2.07]-[2.08]$ by point-splitting the fields in the
internal noncommutative vertices.  Then, phases which combine to
produce planar, phase-independent pieces would be supplemented by
additional, $\epsilon$ dependent phases which survive to regulate the momentum
integrals.  It would be interesting
to carry out this construction, as it might have wider applicability
to noncommutative field theories in general: the standard commutative
divergences that survive as the planar pieces of noncommutative graphs
might be controlled in a way analogous their nonplanar counterparts.

Presumably, the graphs $[1.01]-[1.06]$ are also made rigourously
finite by the extension of the point-splitting regulator.  However, in
order to exhibit the
intricate cancellations that we found amongst them, we
found it necessary to explicitly evaluate their path parameter
integrations, and perform some trigonometric gymnastics with the
unregulated phases.  It is less obvious therefore, that the
cancellations found there will be automatically preserved by careful
application of this regulator.  It is possible that identities
involving $\star_n$ products\cite{Mehen:2000,Liu:2000} might simplify
this computation, since
the complication of having to evaluate these integrals
to find the cancellations arises intrinsically from the higher order expansion
of the Wilson lines.  Finally, the graphs
$[1.07]-[1.10]$ are similar to $[2.07]-[2.08]$\footnote{Certainly it would be
surprising if the cancellation we found between $[1.07]-[1.08]$ and
$[1.09]-[1.10]$ did not hold rigourously, since this directly
generalizes the one-loop cancellation.} in that they also contain the
one-loop vertex correction graphs as subgraphs, and so once the latter are
properly regularized, the former will be too.

Let us now consider the possible generalization of our results to higher
orders.  Clearly, proceeding as we have done to higher orders, even
formally, would quickly become prohibitive.  However, our calculations here
in conjunction with those in \cite{Kaminsky:2003} strongly suggest that
the cancellation of quantum corrections to the correlator holds
at any order in perturbation theory, and that the mechanisms
which enforce such cancellations are already present at the orders we
have studied.  Specifically, we have found that we can organize the 
calculation according to the number of sources on the pure Wilson line
$W(k)$.  Our work suggests that it is sufficient to consider
the simpler correlators at $O(g^{2n-1})$ given by
\be
\vev{ \prod_{i=1}^{m} \kt{1i} A_{\m{1i}}(p_{1i}) O_{\mu\nu}(\kp)}
\delta^{(3)}( \sum_{j=1}^{m} p_{1j}-k ) ~ ,~ m= 1... n.
\ee
Moreover, we have seen that the identity (\ref{master identity}),
\be
 \left[ C^{\rho} C_{[\mu}\epl{\nu]\rho\m{i}} + \epl{\mu\nu\m{i}} C^2\right]
\kt{i} = (k\times C) \epl{\mu\nu\rho} C^{\rho},
\ee
lies at the heart of all of the cancellations, and allows us
to compare graphs with different number of propagators.  The first
term on the left hand side arises from graphs that involve the
derivative term in the field strength, while the second arises from
the commutator term, and is associated with similar graphs with one
less propagator and vertex.  The right hand
side yields a surface term with respect to a path
parameter integration in $O_{\mu\nu}(\kp)$, and is also associated
with a graph involving the 
field strength commutator, but with one more propagator and with one
less source from the path-ordered exponential part of $O_{\mu\nu}(\kp)$.

These observations suggest how to find the cancellations at a
given order $O(g^{2n-1})$.  Consider any graph with $i\le n$ sources on
$W(k)$, $j < n$ sources on the path-ordered exponential component of 
$O_{\mu\nu}(\kp)$ (henceforth denoted $W_2$), and the one source
from the derivative term in the field strength.  Call this
graph I, and fix $i$ for the remainder of this discussion.  To graph I is
always paired a graph also with $(i,j)$ sources on $W(k)$ and
$W_2$ respectively, but with two sources from the field strength
commutator, one less vertex, one less propagator and otherwise
identical to I. Call this graph II.  Colloquially, we can think of
forming it from graph I by collapsing the propagator joining the field
strength derivative term with an internal vertex, into the field strength
insertion, and removing the vertex.  The other two propagators
entering into the internal vertex now connect directly to the field
strength commutator (and the two ways of doing this compensate for the
sinusoidal phase that we lose by removing the vertex).  To
each pair of such graphs, we apply (\ref{master identity}).  In the
case where $j=0$, the right hand side is zero and we are done.
Otherwise build graph III, formed from graph I by 'curling' one of
the sources on $W_2$ into the field strength insertion (see
$[2.01]\rightarrow [2.03]$ for example).  In doing so, we lose a path
parameter integration and source along $W_2$, so $j\rightarrow j-1$, and
contract with the field strength commutator as opposed to the field
strength derivative.  In the $i=1, j=1$ case studied in 
\cite{Kaminsky:2003}, and in the $i=2, j=1$ case in $[2.01]-[2.03]$,
this provides the right hand side of 
(\ref{master identity}), and hence a cancellation with graphs I and II.  

Otherwise we have to consider the
{\it set} of graphs $\{ \tn{I}a, \tn{I}b,... \}$ with fixed $i$ and
$j$, and their partners 
$\{ \tn{II}a, \tn{II}b,... \}$.   We then build all of the graphs III
which can be formed from the I's by curling a source on $W_2$ into the field
strength to obtain a graph with $(i, j-1)$ sources on $W(k)$ and $W_2$
respectively, but with the same number of propagators, as in $[1.05]$
and $[1.06]$.  These will cancel all of the I and II pairs via
(\ref{master identity}), but as we
saw in $[2.06]$ and $[1.06]$, will generally leave residual
terms proportional to $\epl{\mu\nu\m{ij}} \kt{ij}$.  These will occur
in the graphs III where the propagators from the field strength
commutator sources connect to distinct vertices.  To each such graph, we
then invert the
process we used in forming II from I, {\it inserting} a propagator which
joins the derivative term in the field strength to a new vertex, and
creating an internal loop in the process to form a new graph.  Thus,
for example, we
obtain $[2.07]$ and $[1.07]$ from $[2.06]$ and $[1.06]$.  Applying 
(\ref{master identity}) to the terms proportional to 
$C_{[\mu} \epl{\nu] \rh{a} \m{ij}}$ in these new graphs, we
cancel the aforementioned residual pieces, but generate
new surface terms from the right hand side of (\ref{master
identity}), if $C$ depends on a loop momentum, or equivalently if
$j>0$, while the ghost graph cancels
the terms not proportional to $C_{[\mu} \epl{\nu] \rh{a} \m{ij}}$.  Treating
this new graph as one of type I, we then iterate the process of
curling a source on $W_2$ into the field strength insertion, and create a
new graph of type III, thereby reducing $j$ further.  Thus we obtain
$[1.09]$ from $[1.07]$.\footnote{Alternatively note that $[1.07]$ and
$[1.09]$, along with $[1.06]$, are like the graphs we considered in
\cite{Kaminsky:2003}, with one-loop vertex corrections in place of the
vertex itself.  Thus by reducing $j$, we are really setting up a form of
induction.}  This process terminates when $j=0$, since
$C=k$ in (\ref{master identity}) and the right hand side vanishes.

Thus far this discussion has been essentially
independent of the other structure that might be present in the
graphs, because the cancellations we found primarily occur between the
derivative (plus noncommutative Chern-Simons vertex) and commutator
terms in the field strength, in conjunction with the Wilson line
expansion and the noncommutative phases it generates.  Chern-Simons
theory is such that quantum corrections to the correlators of the
basic fields, and the one-particle irreducible functions are
essentially trivial.  While this has been explored in a covariant
gauge at only one-loop in the noncommutative case (see
\cite{Bichl:2000,Chen:2000} for example), and partially at
two-loops here, there is no reason as yet to suspect that the
well-known commutative results at higher orders do not extend to the
noncommutative case.  (See however \cite{Das:2001,Martin:2001} for
axial gauge results involving basic fields in noncommutative 
Chern-Simons theory.)  If we assume this to be true, then
we expect to be able to effectively neglect or reduce the internal
loops in our graphs, and
apply the above arguments to the ones whose loops are formed solely with
the Wilson line noncommutative structure to establish cancellations at
an arbitrary order.

To conclude, we have explicitly shown that the order $g^5$, two-loop
quantum corrections to the correlator of a open Wilson line
$W(k)$, and an open Wilson line with a field strength insertion
$O_{\mu\nu}(\kp)$ in noncommutative Chern-Simons theory cancel amongst
themselves.  This lends further
support to the conjecture that noncommutative and commutative
Chern-Simons theories are perturbatively equivalent.  It would be
interesting both to employ the regulator introduced in
\cite{Kaminsky:2003} and extended above to make the results presented
here rigourous, as well as make explicit the heuristic argument just
discussed to establish an all-orders result at least formally.  We
have focussed on $\vev{W(k)O_{\mu\nu}(\kp)}$ because it is
the simplest nontrivial correlator of composite, gauge-invariant
objects, but it would also be interesting
to examine other correlators at higher orders in the theory, such as
the pure three point function of open Wilson lines $\vev{W(k_1) W(k_2)
W(k_3)}$ we studied at the lowest nontrivial order in
\cite{Kaminsky:2003}.  Here the metric independence of the non-gauge
fixed theory does not forbid nontrivial dependence on $k_1 \times
k_2$, but equivalence with the commutative theory does; a 
preliminary investigation into the correlator  at $O(g^6)$ has not revealed
an obvious set of complete cancellations.  Most importantly however,
the equivalence or inequivalence of the two
theories at a nonperturbative level needs to be established.

%%%%% Acknowledgements %%%%%
\section*{Acknowledgements}
K.K. would like to thank Y. Okawa and H. Ooguri for useful discussions
in the early phase of this project.  This work was supported in part by
the Natural Sciences and Engineering Research Council of Canada.

%\newpage

%%%%%%%%%% Appendices %%%%%%%%%%
\appendix
\renewcommand{\thesection}{Appendix \Alph{section}.}
\renewcommand{\theequation}{\Alph{section}.\arabic{equation}}

%%%%% Appendix A %%%%%
\section{Conventions and Feynman rules}
\setcounter{equation}{0}
The action of noncommutative Chern-Simons theory
in terms of a canonically normalized gauge field is given by
\be
  S_{NCCS} = \frac{1}{2} \int d^3 x~ \epsilon^{\mu \rho \nu}
  \left[
    A_\mu \ast \partial_\rho A_\nu
    - \frac{2ig}{3} A_\mu \ast A_\rho \ast A_\nu
  \right],
\ee
while the standard (noncommutative) ghost action is given by
\be
S_{ghost} = \int d^3 x~ \partial^{\mu}\bar{c}\ast D_{\mu} c,
\ee
where $D_{\mu}(c) \equiv \partial_{\mu} c - ig \left(A_{\mu} \ast c - 
c \ast A_{\mu}\right)$.
Our Fourier transform convention is
\be
  A_\mu (x) = \int \frac{d^3 p}{(2\pi)^3}~
              e^{-ikx} A_\mu (k).
\ee
We use the standard covariant gauge-fixing term proportional to
$(\partial \cdot A)^2$ and then take the Landau gauge, which is known
to be infrared safe, at least in perturbative commutative Chern-Simons theory.

Our path ordering convention (with respect to the star product) for
the open Wilson lines puts larger path parameter values on the left:
\bea
& & P_{\ast}\exp\left[ig\int_0^1 d\sigma \kt{} 
 A_{\mu}\left(x+\xi(\sigma)\right) \right] \non  =  
1 + ig \int_0^1 d\sigma_1 \kt{}A_{\mu}(x+\xi(\sigma_1))
\non \\
& & ~ + (ig)^2 \int_0^1 d\sigma_1 \int_0^{\sigma_1}d\sigma_2~ 
\kt{1}A_{\m{1}}(x+\xi(\sigma_1)) \ast
\kt{2}A_{\m{2}}(x+\xi(\sigma_2)) + O(g^3),
\eea
where $\xi^{\mu}_i= \xi^{\mu}(\sigma_i) = \kt{} \sigma_i \equiv
k_{\nu}\theta^{\nu\mu} \sigma_i$. In the expansion of the pure open
Wilson line, one
of the path integrations at each order is redundant, so that we will
use for example the fact that
\bea
& & \int d^3x \int_0^1 d\sigma_1 \int_0^{\sigma_1} d\sigma_2~
 (k\theta) \cdot A(x+\xi(\sigma_1)) \ast (k\theta) \cdot
A(x+\xi(\sigma_2)) \ast \e{ik\cdot x}  \non \\ & & 
= \frac{1}{2} \int d^3x \int_0^1
d\sigma~ (k\theta)\cdot A(x+\xi(\sigma)) \ast (k\theta)
\cdot A(x) \ast \e{ik\cdot x}.
\label{path simplification}
\eea
Since we always work in momentum space, the following identities,
which use the momentum and index conventions discussed in section 2,
will be useful for our calculations:
\bea
\int d^3x \int_0^1 d\sigma_{11} \int_0^{\sigma_{11}} d\sigma_{21}~
A_{\m{11}}(x+\xi(\sigma_{11}))\ast A_{\m{21}}(x+\xi(\sigma_{21})) \ast \e{i k
\cdot x} & & \non \\
= \frac{1}{2} \int_0^1 d\sigma_1 \int \frac{d^3p_{11} d^3p_{21}}{(2\pi)^3}
\de{p_{11}+p_{21}-k} \e{-i (k \times p_{11}) \sigma_{11}}
\e{-\frac{i}{2} p_{11}\times p_{21}} A_{\m{11}}(p_{11})
A_{\m{2}}(p_{21}),
\eea
\bea
&& \int d^3x \int_0^1 d\sigma_{12} \int_0^{\sigma_{12}} d\sigma_{22}~
A_{\m{12}}(x+\xi(\sigma_{12}))\ast A_{\m{22}}(x+\xi(\sigma_{22})) \ast
\partial_{\mu} A_{\nu}(x) \ast \e{i \kp \cdot x}  \non \\
= &&  \int_0^1 d\sigma_{12} \int_0^{\sigma_{12}} d\sigma_{22} \int 
\frac{d^3p_{12}
d^3p_{22} d^3p_{42}}{(2\pi)^6}~ \de{p_{12}+ p_{22}+ p_{42} - \kp}~ 
\e{- i [(\kp\times p_{12})\sigma_{12} +(\kp\times
p_{22})\sigma_{22}]} \non  \\  
&& \qquad \times \e{-\frac{i}{2}\left[ p_{22}\times (p_{42}-\kp) - p_{42}\times
\kp\right]} (-i p_{42})_{\mu} A_{\m{12}}(p_{12}) A_{\m{22}}(p_{22})
A_{\nu}(p_{42}),
\eea
\bea
&& \int d^3x \int_0^1 d\sigma_{12} \int_0^{\sigma_{12}} d\sigma_{22}~
A_{\m{12}}(x+\xi(\sigma_{12}))\ast A_{\m{22}}(x+\xi(\sigma_{22})) \ast
A_{\mu}(x) \ast A_{\nu}(x) \ast \e{i \kp \cdot x} \non \\
= &&  \int_0^1 d\sigma_{12} \int_0^{\sigma_{12}} d\sigma_{22} \int 
\frac{{\sst d^3p_{12} d^3p_{22}d^3p_{32} d^3p_{42}} }{(2\pi)^9}~ 
\de{\sum_i p_{i2} - \kp}~ 
\e{- i [(\kp\times p_{12})\sigma_{12} +(\kp\times
p_{22})\sigma_{22}]} \non  \\  
&& \qquad \times \e{-\frac{i}{2}\left[ p_{22}\times (p_{32}+p_{42}-\kp)+
p_{32} \times (p_{42}-\kp) - p_{42}\times
\kp\right]}  A_{\m{12}}(p_{12}) A_{\m{22}}(p_{22}) A_{\mu}(p_{32}) A_{\nu}(p_{42}),
\eea
where $k\times p \equiv k\theta \cdot p = \tilde{k} \cdot p = 
k_{\mu}\theta^{\mu\nu} p_{\nu}$.

The momentum space Feynman rules for the gauge field and ghost
propagators respectively are
\bea
 p, \mu \epsfxsize=1in\epsfbox{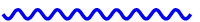} q, \nu
  & = &  (2 \pi)^3 \De{p+q} \epl{\mu \rho \nu}
    \frac{p^\rho}{p^2} \\
 p \hspace{0.1cm} \epsfxsize=1in\epsfbox{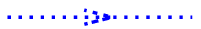} \hspace{0.1cm} q 
    \hspace{0.3cm}
  & = & (2 \pi)^3 \De{p+q} \frac{i}{p^2},
\label{propagators}
\eea
while the Feynman rules for the triple gauge, and the
ghost-antighost-gauge vertices are given respectively by
\bea
 \q2,\al2 \hspace{1.3cm} & & \nonumber \\
 \epsfxsize=1in\epsfbox{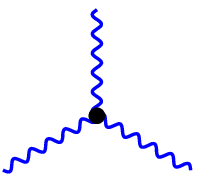} \hspace{0.4cm} 
    & = & -2ig \De{\sum_i q_i} \sin\left( 
   \frac{\q1\times\q2}{2} \right) \epu{\al1\al2\al3} \\
    \q1, \al1 \hspace{1.3cm} \q3, \al3  & & \nonumber \\
  \mu, \q2 \hspace{1.3cm} & & \nonumber \\
 \epsfxsize=1in\epsfbox{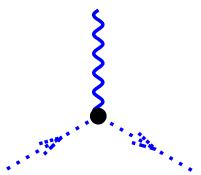}\hspace{0.4cm}
     & = & -2ig \De{\sum_i q_i} \sin\left( 
   \frac{\q1\times\q2}{2} \right) (-i\q1)^{\mu}, ~ \\
  \q1  \hspace{2.1cm} \q3  \hspace{0.2cm} & & \nonumber
 \label{vertices}
\eea
where the momentum $\q1$ appearing in the latter rule is associated 
with the antighost.

We will extensively use the following identities involving the
antisymmetric tensor:
\bea
\epu{a AB}\epl{a CD}& = & \delta^A_C \delta^B_D - \delta^A_D
\delta^B_C \equiv \delta^{AB}_{CD} - \delta^{AB}_{DC} \label{epsilon
identity 1} \\
\epl{ABC}\epl{DEF}& = & \epl{ABD}\epl{CEF}+\epl{ABE}\epl{CFD}+ 
\epl{ABF}\epl{CDE} \label{epsilon identity 2},
\eea
the second of which yields
\be
\epl{\rh{a}\rh{b}[\mu}\epl{\nu]\rh{c}\m{ij}}  =
\epl{\mu\nu\rh{c}}\epl{\m{ij}\rh{a}\rh{b}} - 
\epl{\mu\nu\m{ij}}\epl{\rh{a}\rh{b}\rh{c}}.  
\label{epsilon identity 3}
\ee
Contracting with two factors of $\kt{i}$ gives
\be
\epl{\mu\rh{a}\m{i}}\epl{\nu\rh{c}\m{j}} \kt{1}\kt{2} -
(\mu\leftrightarrow\nu) = \epl{\mu\nu\m{i}}\epl{\m{j}\rh{a}\rh{c}}
\kt{i}\kt{j}.
\label{epsilon identity 4}
\ee
Finally, the central identity that underlies most of the
cancellations between Feynman diagrams we will exhibit is given by
\be
 \left[ C^{\rho} C_{[\mu}\epl{\nu]\rho\m{i}} + \epl{\mu\nu\m{i}} C^2\right]
\kt{i} = (k\times C) \epl{\mu\nu\rho} C^{\rho}.
\label{master identity}
\ee

%%%%%%%%%% References %%%%%%%%%%

\renewcommand{\baselinestretch}{0.87}

%\bibliography{draft}
%\bibliographystyle{ssg}
\begingroup\raggedright\endgroup
\end{document}